\documentclass[reqno]{amsart}%

\usepackage{graphicx}
\usepackage{amsmath}
\usepackage{amssymb}
\usepackage{amsbsy}

\usepackage{bm}
\bmdefine{\va}{\alpha}
\newcommand{\dlift}[1]{{\mathbf L}_{#1}}
\newcommand{\idlift}[1]{{\mathbf L}_{#1}^{-1}}

\newcommand{\sspl}[4]{\left[\begin{array}{c|c}#1&#2\\\hline#3&#4\end{array}\right]}
\newcommand{\ctd}[2]{{\mathcal{D}}_{#1}\left(#2\right)}
\newcommand{\lift}[2]{{\mathcal{L}}_{#1}\left(#2\right)}

\newcommand{\hold}[1]{{\mathcal H}_{#1}}
\newcommand{\samp}[1]{{\mathcal S}_{#1}}
\newcommand{\us}[1]{\uparrow\!#1}

\newcommand{\real}{{\mathbb R}}

\newcommand{\M}{{\mathbf M}}
\newcommand{\K}{{\mathcal K}}
\newcommand{\tK}{\widetilde{\K}}
\newcommand{\ds}[1]{\downarrow\!#1}

\newcommand{\BL}{BL}
\newcommand{\FL}{{FL^2}}
\newcommand{\jj}{{\mathrm{j}}}
\newcommand{\dd}{{\mathrm{d}}}
\newcommand{\supp}{\mathrm{supp}}

\newtheorem{prob}{Problem}
\newtheorem{lemma}{Lemma}

\newtheorem{theorem}{Theorem}

\begin{document}

\title[$H^\infty$ Design of Periodically Nonuniform Interpolation and Decimation]
{$H^\infty$ Design of Periodically Nonuniform Interpolation and Decimation
for Non-Band-Limited Signals}
\author[M. Nagahara]{Masaaki Nagahara}
\author[M. Ogura]{Masaki Ogura}
\author[Y. Yamamoto]{Yutaka Yamamoto}
\address{
M. Nagahara and Y. Yamamoto are with Graduate School of Informatics, Kyoto University.
M. Ogura is with Department of Mathematics and Statistics,
Texas Tech University.
The corresponding author is M. Nagahara (nagahara@ieee.org).
}

\begin{abstract}%
In this paper, we consider signal interpolation
of discrete-time signals which are decimated nonuniformly.
A conventional interpolation method is based on the  sampling theorem,
and the resulting system consists of an ideal filter with complex-valued coefficients.
While the conventional method assumes band limitation of signals,
we propose a new method by sampled-data $H^\infty$ optimization.
By this method, we can remove the band-limiting assumption and the optimal filter
can be with real-valued coefficients.
Moreover, we show that without band-limited assumption, there can be the optimal 
decimation patterns among ones with the same ratio.
By examples, we show the effectiveness of our method.
\end{abstract}

\maketitle
\section{Introduction}
Interpolation is a fundamental operation in digital signal processing,
and has many applications such as signal reconstruction,
signal compression/expansion, and resizing/rotating digital images
\cite{Vai,KleZam}.
%see \cite{Vai}
If digital data (discrete-time signals) to be interpolated are spaced uniformly on the time axis,
the {\it uniform} interpolation is executed by an expander and a digital filter 
(called an interpolation filter) \cite{Vai},
which is conventionally designed via 
the sampling theorem.

%On the other hand,
{\it Periodically nonuniform} interpolation (or decimation) 
also plays an important role in signal processing,
such as signal compression by nonuniform filterbanks \cite{NayBarSmi93},
super-resolution image processing \cite{ParParKan03}, and
time-interleaved AD converters \cite{StrTan07}.
The design has been studied by many researchers 
\cite{VaiLiu90,NayBarSmi93,VanDelMaiCuv97,LinVai98,MarNar02},
in which the design methods are  based on the generalized sampling theorem,
assuming that the original signals to be sampled are {\it band-limited}
below the Nyquist frequency.  
Then the optimal filter (or the perfect reconstruction filter)
is given by an ideal lowpass filter with complex coefficients
\cite{VaiLiu90,Vai}.
Since the ideal filter cannot be realized,
approximation methods are also proposed; see in particular \cite{VaiLiu90,VanDelMaiCuv97}.

On the other hand, real signals such as audio signals (esp.~orchestral music)
violate the band-limiting assumption in the sampling theorem,
that is, they have some frequency components beyond the Nyquist frequency.
In view of this, we have to take account of the {\it whole} frequency range
in designing interpolation systems.
For this purpose,
{\it sampled-data $H^\infty$ optimization} \cite{CheFra,NagYam00}
is very adequate.

A similar philosophy has also been presented and proposed
by Unser and co-workers \cite{UnsZer98,Uns05}.
This method is a generalization of Shannon sampling theory 
intending to give a machinery that works for signals that 
are not necessarily perfectly band-limited.  The method works
well for those analog signals that belong to a prespecified subspace,
but not necessarily so for those that do not.  It is even shown 
that their method can lead to 
an unstable reconstruction filter \cite{NagYamKha08}.  
Moreover, the signal subspace is constructed 
by the linear span of a given generating function, but 
it is not easy to identify the generating function 
in real applications.  
On the other hand, our approach models the signal subspace
in terms of analog (continuous-time) frequency characteristics,
which can easily be identified by modeling a signal generator
by physical laws or through the Fourier transform of real signals.

The main objective of this paper is to propose a new design method
for nonuniform interpolation via sampled-data $H^\infty$ optimization.
This design problem is formulated by minimizing the $H^\infty$ norm ($L^2$-induced norm)
of the error system between the delayed original analog signals
and the output of the reconstruction system.
Since this error system includes both continuous- and discrete-time signals (systems),
the optimization is an infinite dimensional one.
To convert this to a finite dimensional optimization,
we introduce the fast discretization method 
\cite[Chap.~8]{CheFra},~\cite{YamMadAnd99}.  
By this method, the optimal interpolation filter can be obtained by 
numerical computations.
MATLAB codes for this optimization are available through \cite{MAT}.

We also show in this paper that there are cases with the 
same decimation rates but 
the optimal reconstruction performances can differ when the 
decimation patterns are different.
That is, the performance depends on the decimation pattern.
Note that this property cannot be captured via the sampling theorem.

The paper is organized as follows. We first define nonuniform decimation with a decimation pattern
in Section \ref{sec:nonuniform-decimation}.
We show that this definition includes the block decimation introduced in \cite{NayBarSmi93}.
In Section \ref{sec:design} we define nonuniform expanders and formulate the interpolation problem using such an expander for non-band-limited signals.
A design procedure and implementation as a multirate filterbank are also given in this section.
In Section \ref{sec:optimal-pattern},
we consider optimal decimation pattern analysis.
Section \ref{sec:examples} shows design examples.
In this section, we will show a result of our optimization and compare it with
a conventional design proposed in \cite{VaiLiu90,Vai}.
Here we will show our method is superior to the conventional method.
Optimal decimation patterns for several decimation ratios are also 
presented.  
Section \ref{sec:conclusion} concludes our result.

\section*{Notation}
Throughout this paper, we use the following notation:
\begin{description}
\item[$\real$, $\real^{M}, \real^{M\times N}$] the sets of real numbers, real valued column vectors
of size $M$, and $M$ by $N$ real valued matrices, respectively.
\item[$L^2$] the Lebesgue space consisting of all square integrable real functions.
\item[$z$] the symbol for $Z$ transform (the $Z$ transform of the forward shift operator).
\item[$s$] the symbol for Laplace transform (the Laplace transform of the differentiator $\frac{\dd}{\dd t}$).
\item[$A^\top$] the transpose of a matrix $A$.
\item[$I_{N}$] the $N\times N$ identity matrix.
\item[$\mathbf{blockdiag}(A,B,\ldots,C)$] a block-diagonal matrix of matrices $A$, $B$, $\ldots$, $C$, that is,
\[
 \mathbf{blockdiag}(A,B,\ldots,C) = \left[\begin{array}{cccc}A& & & \\ &B& & \\ & &\ddots& \\ & & &C\end{array}\right].
\]
\item[$\sspl{A}{B}{C}{D}$] a state-space representation for a continuous-time system
$C(sI-A)^{-1}B+D$  or discrete-time one $C(zI-A)^{-1}B+D$.

\end{description}
\section{Nonuniform Decimation and Decimation Patterns}
\label{sec:nonuniform-decimation}
Let us consider the discrete-time signal 
$
x:=\{x_0, x_1, x_2, \ldots \}.
$
%shown in Fig.\ \ref{fig:sampled-data}.
Then nonuniform decimation by $\M:=[1,1,0]$ 
(we call this a {\it decimation pattern}) 
is defined by
\begin{equation}
\label{eq:decimated}
(\ds{\M})x := \{x_0, x_1, x_3, x_4, x_6,\ldots\}.
\end{equation}
That is, we first divide the time axis into segments of length three
(the number of the elements in $\M$),
then, in each segment, 
retain the samples corresponding to 1 in $\M$ and discard the samples corresponding to 0.
We then define a general nonuniform decimation with decimation pattern
\begin{equation}
\M := \left[\begin{array}{cccc} b_0 & b_1 & \ldots & b_{M-1} \end{array}\right],\quad b_i \in \{0,1\},
%\M := [b_0, b_1,\ldots, b_{M-1}],\quad b_i \in \{0,1\},
\label{eq:pattern}
\end{equation}
where $M$ is the number of elements in $\M$.
Let $i_1,i_2,\ldots,i_N$ satisfying
\[
0\leq i_1<i_2<\ldots<i_N \leq M-1
\]
be the indices
of $b_i$'s such that
$b_{i_1}=\cdots=b_{i_N}=1$
where $N$ is the number of ones in $\M$.
Then for $x=\{x_0,x_1,\ldots\}$, the nonuniform decimation is defined by
\[
 (\ds{\M})x = \left\{x_{i_1},x_{i_2},\ldots,x_{i_N},x_{M+i_1},x_{M+i_2},\ldots\right\}.
\]
This definition includes the so-called block decimation \cite{NayBarSmi93}, 
in which the first $R_1$ samples of each segment of $R_2$ samples are 
retained while the rest are discarded.
By our notation,
the block decimation $R_2:R_1$ is represented as $\ds{\M}$ with
\[
\M={[\underbrace{1,\ldots,1}_{R_1},\underbrace{0,\ldots,0}_{R_2-R_1}]}.
\]

The {\it decimation ratio of $\M$} is defined to be $M/N$.
Note that since $N\leq M$, the ratio is always greater than or equal to 1.
By our definition,
the uniform decimator $\ds{M}$ where $M$ is a positive integer 
is represented as a special case of nonuniform decimator
\[
\ds{M} = \ds{\M},\quad \M = [1, \underbrace{0,0,\ldots,0}_{M-1}],
\]
with decimation ratio $M$.

\section{Design of Interpolation Filter}
\label{sec:design}
\subsection{Nonuniform Interpolation}
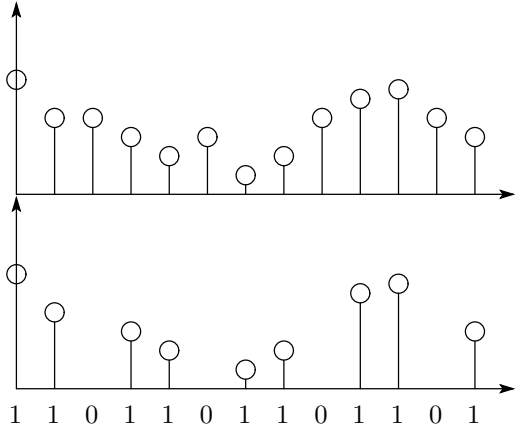
\begin{figure}[tb]
\begin{center}
%%%%\input{sampled-data.tex}
%WinTpicVersion3.08
\unitlength 0.1in
\begin{picture}( 26.5000, 10.0500)(  3.5000,-12.0500)
% VECTOR 2 0 3 0
% 2 400 1200 400 200
% 
\special{pn 8}%
\special{pa 400 1200}%
\special{pa 400 200}%
\special{fp}%
\special{sh 1}%
\special{pa 400 200}%
\special{pa 380 268}%
\special{pa 400 254}%
\special{pa 420 268}%
\special{pa 400 200}%
\special{fp}%
% VECTOR 2 0 3 0
% 2 400 1200 3000 1200
% 
\special{pn 8}%
\special{pa 400 1200}%
\special{pa 3000 1200}%
\special{fp}%
\special{sh 1}%
\special{pa 3000 1200}%
\special{pa 2934 1180}%
\special{pa 2948 1200}%
\special{pa 2934 1220}%
\special{pa 3000 1200}%
\special{fp}%
% CIRCLE 2 0 3 0
% 4 400 600 400 650 400 650 400 650
% 
\special{pn 8}%
\special{ar 400 600 50 50  0.0000000 6.2831853}%
% CIRCLE 2 0 3 0
% 4 600 800 600 850 600 850 600 850
% 
\special{pn 8}%
\special{ar 600 800 50 50  0.0000000 6.2831853}%
% CIRCLE 2 0 3 0
% 4 800 800 800 850 800 850 800 850
% 
\special{pn 8}%
\special{ar 800 800 50 50  0.0000000 6.2831853}%
% CIRCLE 2 0 3 0
% 4 1000 900 1000 950 1000 950 1000 950
% 
\special{pn 8}%
\special{ar 1000 900 50 50  0.0000000 6.2831853}%
% CIRCLE 2 0 3 0
% 4 1200 1000 1200 1050 1200 1050 1200 1050
% 
\special{pn 8}%
\special{ar 1200 1000 50 50  0.0000000 6.2831853}%
% CIRCLE 2 0 3 0
% 4 1400 900 1400 950 1400 950 1400 950
% 
\special{pn 8}%
\special{ar 1400 900 50 50  0.0000000 6.2831853}%
% CIRCLE 2 0 3 0
% 4 1600 1100 1600 1150 1600 1150 1600 1150
% 
\special{pn 8}%
\special{ar 1600 1100 50 50  0.0000000 6.2831853}%
% CIRCLE 2 0 3 0
% 4 1800 1000 1800 1050 1800 1050 1800 1050
% 
\special{pn 8}%
\special{ar 1800 1000 50 50  0.0000000 6.2831853}%
% CIRCLE 2 0 3 0
% 4 2000 800 2000 850 2000 850 2000 850
% 
\special{pn 8}%
\special{ar 2000 800 50 50  0.0000000 6.2831853}%
% CIRCLE 2 0 3 0
% 4 2200 700 2200 750 2200 750 2200 750
% 
\special{pn 8}%
\special{ar 2200 700 50 50  0.0000000 6.2831853}%
% CIRCLE 2 0 3 0
% 4 2400 650 2400 700 2400 700 2400 700
% 
\special{pn 8}%
\special{ar 2400 650 50 50  0.0000000 6.2831853}%
% CIRCLE 2 0 3 0
% 4 2600 800 2600 850 2600 850 2600 850
% 
\special{pn 8}%
\special{ar 2600 800 50 50  0.0000000 6.2831853}%
% CIRCLE 2 0 3 0
% 4 2800 900 2800 950 2800 950 2800 950
% 
\special{pn 8}%
\special{ar 2800 900 50 50  0.0000000 6.2831853}%
% LINE 2 0 3 0
% 2 600 1200 600 850
% 
\special{pn 8}%
\special{pa 600 1200}%
\special{pa 600 850}%
\special{fp}%
% LINE 2 0 3 0
% 2 800 1200 800 850
% 
\special{pn 8}%
\special{pa 800 1200}%
\special{pa 800 850}%
\special{fp}%
% LINE 2 0 3 0
% 2 1000 1200 1000 950
% 
\special{pn 8}%
\special{pa 1000 1200}%
\special{pa 1000 950}%
\special{fp}%
% LINE 2 0 3 0
% 2 1200 1200 1200 1050
% 
\special{pn 8}%
\special{pa 1200 1200}%
\special{pa 1200 1050}%
\special{fp}%
% LINE 2 0 3 0
% 2 1400 1200 1400 950
% 
\special{pn 8}%
\special{pa 1400 1200}%
\special{pa 1400 950}%
\special{fp}%
% LINE 2 0 3 0
% 2 1600 1150 1600 1200
% 
\special{pn 8}%
\special{pa 1600 1150}%
\special{pa 1600 1200}%
\special{fp}%
% LINE 2 0 3 0
% 2 1800 1050 1800 1200
% 
\special{pn 8}%
\special{pa 1800 1050}%
\special{pa 1800 1200}%
\special{fp}%
% LINE 2 0 3 0
% 2 2000 1200 2000 850
% 
\special{pn 8}%
\special{pa 2000 1200}%
\special{pa 2000 850}%
\special{fp}%
% LINE 2 0 3 0
% 2 2200 1200 2200 750
% 
\special{pn 8}%
\special{pa 2200 1200}%
\special{pa 2200 750}%
\special{fp}%
% LINE 2 0 3 0
% 2 2400 1200 2400 700
% 
\special{pn 8}%
\special{pa 2400 1200}%
\special{pa 2400 700}%
\special{fp}%
% LINE 2 0 3 0
% 2 2600 1200 2600 850
% 
\special{pn 8}%
\special{pa 2600 1200}%
\special{pa 2600 850}%
\special{fp}%
% LINE 2 0 3 0
% 2 2800 1200 2800 950
% 
\special{pn 8}%
\special{pa 2800 1200}%
\special{pa 2800 950}%
\special{fp}%
\end{picture}\\%
%%%%\input{decimated.tex}
%WinTpicVersion3.08
\unitlength 0.1in
\begin{picture}( 26.5000, 10.5000)(  3.5000,-12.5000)
% VECTOR 2 0 3 0
% 2 400 1200 400 200
% 
\special{pn 8}%
\special{pa 400 1200}%
\special{pa 400 200}%
\special{fp}%
\special{sh 1}%
\special{pa 400 200}%
\special{pa 380 268}%
\special{pa 400 254}%
\special{pa 420 268}%
\special{pa 400 200}%
\special{fp}%
% VECTOR 2 0 3 0
% 2 400 1200 3000 1200
% 
\special{pn 8}%
\special{pa 400 1200}%
\special{pa 3000 1200}%
\special{fp}%
\special{sh 1}%
\special{pa 3000 1200}%
\special{pa 2934 1180}%
\special{pa 2948 1200}%
\special{pa 2934 1220}%
\special{pa 3000 1200}%
\special{fp}%
% CIRCLE 2 0 3 0
% 4 400 600 400 650 400 650 400 650
% 
\special{pn 8}%
\special{ar 400 600 50 50  0.0000000 6.2831853}%
% CIRCLE 2 0 3 0
% 4 600 800 600 850 600 850 600 850
% 
\special{pn 8}%
\special{ar 600 800 50 50  0.0000000 6.2831853}%
% CIRCLE 2 0 3 0
% 4 1000 900 1000 950 1000 950 1000 950
% 
\special{pn 8}%
\special{ar 1000 900 50 50  0.0000000 6.2831853}%
% LINE 2 0 3 0
% 2 1200 1200 1200 1050
% 
\special{pn 8}%
\special{pa 1200 1200}%
\special{pa 1200 1050}%
\special{fp}%
% CIRCLE 2 0 3 0
% 4 1200 1000 1200 1050 1200 1050 1200 1050
% 
\special{pn 8}%
\special{ar 1200 1000 50 50  0.0000000 6.2831853}%
% CIRCLE 2 0 3 0
% 4 1600 1100 1600 1150 1600 1150 1600 1150
% 
\special{pn 8}%
\special{ar 1600 1100 50 50  0.0000000 6.2831853}%
% CIRCLE 2 0 3 0
% 4 1800 1000 1800 1050 1800 1050 1800 1050
% 
\special{pn 8}%
\special{ar 1800 1000 50 50  0.0000000 6.2831853}%
% CIRCLE 2 0 3 0
% 4 2200 700 2200 750 2200 750 2200 750
% 
\special{pn 8}%
\special{ar 2200 700 50 50  0.0000000 6.2831853}%
% CIRCLE 2 0 3 0
% 4 2400 650 2400 700 2400 700 2400 700
% 
\special{pn 8}%
\special{ar 2400 650 50 50  0.0000000 6.2831853}%
% CIRCLE 2 0 3 0
% 4 2800 900 2800 950 2800 950 2800 950
% 
\special{pn 8}%
\special{ar 2800 900 50 50  0.0000000 6.2831853}%
% LINE 2 0 3 0
% 2 600 1200 600 850
% 
\special{pn 8}%
\special{pa 600 1200}%
\special{pa 600 850}%
\special{fp}%
% LINE 2 0 3 0
% 2 1000 1200 1000 950
% 
\special{pn 8}%
\special{pa 1000 1200}%
\special{pa 1000 950}%
\special{fp}%
% LINE 2 0 3 0
% 2 1600 1150 1600 1200
% 
\special{pn 8}%
\special{pa 1600 1150}%
\special{pa 1600 1200}%
\special{fp}%
% LINE 2 0 3 0
% 2 1800 1050 1800 1200
% 
\special{pn 8}%
\special{pa 1800 1050}%
\special{pa 1800 1200}%
\special{fp}%
% LINE 2 0 3 0
% 2 2200 1200 2200 750
% 
\special{pn 8}%
\special{pa 2200 1200}%
\special{pa 2200 750}%
\special{fp}%
% LINE 2 0 3 0
% 2 2400 1200 2400 700
% 
\special{pn 8}%
\special{pa 2400 1200}%
\special{pa 2400 700}%
\special{fp}%
% LINE 2 0 3 0
% 2 2800 1200 2800 950
% 
\special{pn 8}%
\special{pa 2800 1200}%
\special{pa 2800 950}%
\special{fp}%
% STR 2 0 3 0
% 3 395 1235 395 1335 5 0
% 1
\put(3.9500,-13.3500){\makebox(0,0){1}}%
% STR 2 0 3 0
% 3 595 1235 595 1335 5 0
% 1
\put(5.9500,-13.3500){\makebox(0,0){1}}%
% STR 2 0 3 0
% 3 795 1235 795 1335 5 0
% 0
\put(7.9500,-13.3500){\makebox(0,0){0}}%
% STR 2 0 3 0
% 3 995 1235 995 1335 5 0
% 1
\put(9.9500,-13.3500){\makebox(0,0){1}}%
% STR 2 0 3 0
% 3 1195 1235 1195 1335 5 0
% 1
\put(11.9500,-13.3500){\makebox(0,0){1}}%
% STR 2 0 3 0
% 3 1395 1235 1395 1335 5 0
% 0
\put(13.9500,-13.3500){\makebox(0,0){0}}%
% STR 2 0 3 0
% 3 1595 1235 1595 1335 5 0
% 1
\put(15.9500,-13.3500){\makebox(0,0){1}}%
% STR 2 0 3 0
% 3 1795 1235 1795 1335 5 0
% 1
\put(17.9500,-13.3500){\makebox(0,0){1}}%
% STR 2 0 3 0
% 3 1995 1235 1995 1335 5 0
% 0
\put(19.9500,-13.3500){\makebox(0,0){0}}%
% STR 2 0 3 0
% 3 2195 1235 2195 1335 5 0
% 1
\put(21.9500,-13.3500){\makebox(0,0){1}}%
% STR 2 0 3 0
% 3 2395 1235 2395 1335 5 0
% 1
\put(23.9500,-13.3500){\makebox(0,0){1}}%
% STR 2 0 3 0
% 3 2595 1235 2595 1335 5 0
% 0
\put(25.9500,-13.3500){\makebox(0,0){0}}%
% STR 2 0 3 0
% 3 2795 1235 2795 1335 5 0
% 1
\put(27.9500,-13.3500){\makebox(0,0){1}}%
\end{picture}%
\end{center}
\vspace{5mm}
\caption{Nonuniform decimation and expansion ($\M=[1,1,0]$):
original sequence $x$ (above) and $(\us{\M})(\ds{\M})x$ (below).}
\label{fig:decimated}
\end{figure}

To consider signal reconstruction from a nonuniformly decimated signal
$(\ds{\M})x$, we define the nonuniform expander $\us{\M}$.
Let $\M = [1,1,0]$.
Then we define $(\us{\M})x$ for $x=\{x_0,x_1,\ldots\}$ by
\[
(\us{\M}) x := \{x_0,x_1,0,x_2, x_3, 0, x_4, \ldots\}.
\]
That is, we first divide the time axis into segments of length two
(the number of 1's in $\M$),
then insert 0's into the position corresponding to 0's in $\M$.
By this definition,
the uniform expander $\us{M}$ where $M$ is a positive integer is represented as
a nonuniform expander
\[
\us{M} = \us{\M},\quad \M = [1, \underbrace{0,0,\ldots,0}_{M-1}].
\]
Applying this to the decimated sequence (\ref{eq:decimated}), we have
\[
%\begin{equation}
%\label{eq:interpolation}
v:=(\us{\M}) (\ds{\M}) x = \{x_0,x_1,0,x_3, x_4, 0, x_6, \ldots\}.
%\end{equation}
\]
The procedure is shown in Fig.\ \ref{fig:decimated}.
In general case of the decimation pattern (\ref{eq:pattern}),
the expansion is given by
\[
\begin{split}
 v &= (\us{\M})(\ds{\M}) x \\
   &=
 \{b_0x_0,b_1x_1,\ldots,b_{M-1}x_{M-1},b_0x_{M},b_1x_{M+1},\ldots\}.
\end{split}
\]
\subsection{Polyphase Representation}
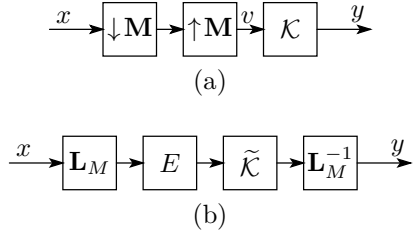
\begin{figure}[tb]
\begin{center}
%%%%\input{interpolation.tex}
%WinTpicVersion3.08
\unitlength 0.1in
\begin{picture}( 21.7500, 11.0000)(  0.2500,-12.2900)
% VECTOR 2 0 3 0
% 2 310 334 590 334
% 
\special{pn 8}%
\special{pa 310 334}%
\special{pa 590 334}%
\special{fp}%
\special{sh 1}%
\special{pa 590 334}%
\special{pa 524 314}%
\special{pa 538 334}%
\special{pa 524 354}%
\special{pa 590 334}%
\special{fp}%
% BOX 2 0 3 0
% 2 590 194 870 474
% 
\special{pn 8}%
\special{pa 590 194}%
\special{pa 870 194}%
\special{pa 870 474}%
\special{pa 590 474}%
\special{pa 590 194}%
\special{fp}%
% VECTOR 2 0 3 0
% 2 870 334 1010 334
% 
\special{pn 8}%
\special{pa 870 334}%
\special{pa 1010 334}%
\special{fp}%
\special{sh 1}%
\special{pa 1010 334}%
\special{pa 944 314}%
\special{pa 958 334}%
\special{pa 944 354}%
\special{pa 1010 334}%
\special{fp}%
% BOX 2 0 3 0
% 2 1010 194 1290 474
% 
\special{pn 8}%
\special{pa 1010 194}%
\special{pa 1290 194}%
\special{pa 1290 474}%
\special{pa 1010 474}%
\special{pa 1010 194}%
\special{fp}%
% VECTOR 2 0 3 0
% 2 1290 334 1430 334
% 
\special{pn 8}%
\special{pa 1290 334}%
\special{pa 1430 334}%
\special{fp}%
\special{sh 1}%
\special{pa 1430 334}%
\special{pa 1364 314}%
\special{pa 1378 334}%
\special{pa 1364 354}%
\special{pa 1430 334}%
\special{fp}%
% BOX 2 0 3 0
% 2 1430 194 1710 474
% 
\special{pn 8}%
\special{pa 1430 194}%
\special{pa 1710 194}%
\special{pa 1710 474}%
\special{pa 1430 474}%
\special{pa 1430 194}%
\special{fp}%
% VECTOR 2 0 3 0
% 2 1710 334 1990 334
% 
\special{pn 8}%
\special{pa 1710 334}%
\special{pa 1990 334}%
\special{fp}%
\special{sh 1}%
\special{pa 1990 334}%
\special{pa 1924 314}%
\special{pa 1938 334}%
\special{pa 1924 354}%
\special{pa 1990 334}%
\special{fp}%
% STR 2 0 3 0
% 3 730 264 730 334 5 0
% $\ds{\M}$
\put(7.3000,-3.3400){\makebox(0,0){$\ds{\M}$}}%
% STR 2 0 3 0
% 3 1150 264 1150 334 5 0
% $\us{\M}$
\put(11.5000,-3.3400){\makebox(0,0){$\us{\M}$}}%
% STR 2 0 3 0
% 3 1570 264 1570 334 5 0
% $\K$
\put(15.7000,-3.3400){\makebox(0,0){$\K$}}%
% STR 2 0 3 0
% 3 345 229 345 299 2 0
% $x$
\put(3.4500,-2.9900){\makebox(0,0)[lb]{$x$}}%
% STR 2 0 3 0
% 3 1885 229 1885 299 2 0
% $y$
\put(18.8500,-2.9900){\makebox(0,0)[lb]{$y$}}%
% $v$
\put(13.1000,-2.9900){\makebox(0,0)[lb]{$v$}}%
% STR 2 0 3 0
% 3 1150 544 1150 614 5 0
% (a)
\put(11.5000,-6.1400){\makebox(0,0){(a)}}%
% BOX 2 0 3 0
% 2 380 894 660 1174
% 
\special{pn 8}%
\special{pa 380 894}%
\special{pa 660 894}%
\special{pa 660 1174}%
\special{pa 380 1174}%
\special{pa 380 894}%
\special{fp}%
% VECTOR 2 0 3 0
% 2 660 1034 800 1034
% 
\special{pn 8}%
\special{pa 660 1034}%
\special{pa 800 1034}%
\special{fp}%
\special{sh 1}%
\special{pa 800 1034}%
\special{pa 734 1014}%
\special{pa 748 1034}%
\special{pa 734 1054}%
\special{pa 800 1034}%
\special{fp}%
% BOX 2 0 3 0
% 2 800 894 1080 1174
% 
\special{pn 8}%
\special{pa 800 894}%
\special{pa 1080 894}%
\special{pa 1080 1174}%
\special{pa 800 1174}%
\special{pa 800 894}%
\special{fp}%
% VECTOR 2 0 3 0
% 2 1080 1034 1220 1034
% 
\special{pn 8}%
\special{pa 1080 1034}%
\special{pa 1220 1034}%
\special{fp}%
\special{sh 1}%
\special{pa 1220 1034}%
\special{pa 1154 1014}%
\special{pa 1168 1034}%
\special{pa 1154 1054}%
\special{pa 1220 1034}%
\special{fp}%
% BOX 2 0 3 0
% 2 1220 894 1500 1174
% 
\special{pn 8}%
\special{pa 1220 894}%
\special{pa 1500 894}%
\special{pa 1500 1174}%
\special{pa 1220 1174}%
\special{pa 1220 894}%
\special{fp}%
% VECTOR 2 0 3 0
% 2 1500 1034 1640 1034
% 
\special{pn 8}%
\special{pa 1500 1034}%
\special{pa 1640 1034}%
\special{fp}%
\special{sh 1}%
\special{pa 1640 1034}%
\special{pa 1574 1014}%
\special{pa 1588 1034}%
\special{pa 1574 1054}%
\special{pa 1640 1034}%
\special{fp}%
% BOX 2 0 3 0
% 2 1640 894 1920 1174
% 
\special{pn 8}%
\special{pa 1640 894}%
\special{pa 1920 894}%
\special{pa 1920 1174}%
\special{pa 1640 1174}%
\special{pa 1640 894}%
\special{fp}%
% VECTOR 2 0 3 0
% 2 1920 1034 2200 1034
% 
\special{pn 8}%
\special{pa 1920 1034}%
\special{pa 2200 1034}%
\special{fp}%
\special{sh 1}%
\special{pa 2200 1034}%
\special{pa 2134 1014}%
\special{pa 2148 1034}%
\special{pa 2134 1054}%
\special{pa 2200 1034}%
\special{fp}%
% VECTOR 2 0 3 0
% 2 100 1034 380 1034
% 
\special{pn 8}%
\special{pa 100 1034}%
\special{pa 380 1034}%
\special{fp}%
\special{sh 1}%
\special{pa 380 1034}%
\special{pa 314 1014}%
\special{pa 328 1034}%
\special{pa 314 1054}%
\special{pa 380 1034}%
\special{fp}%
% STR 2 0 3 0
% 3 520 964 520 1034 5 0
% $\dlift{M}$
\put(5.2000,-10.3400){\makebox(0,0){$\dlift{M}$}}%
% STR 2 0 3 0
% 3 940 964 940 1034 5 0
% $E$
\put(9.4000,-10.3400){\makebox(0,0){$E$}}%
% STR 2 0 3 0
% 3 1360 964 1360 1034 5 0
% $\tK$
\put(13.6000,-10.3400){\makebox(0,0){$\tK$}}%
% STR 2 0 3 0
% 3 1780 964 1780 1034 5 0
% $\idlift{M}$
\put(17.8000,-10.3400){\makebox(0,0){$\idlift{M}$}}%
% STR 2 0 3 0
% 3 135 929 135 999 2 0
% $x$
\put(1.3500,-9.9900){\makebox(0,0)[lb]{$x$}}%
% STR 2 0 3 0
% 3 2095 929 2095 999 2 0
% $y$
\put(20.9500,-9.9900){\makebox(0,0)[lb]{$y$}}%
% STR 2 0 3 0
% 3 1150 1244 1150 1314 5 0
% (b)
\put(11.5000,-13.1400){\makebox(0,0){(b)}}%
\end{picture}%

\vspace{5mm}
\end{center}
\caption{(a) Nonuniform decimation and interpolation,  (b) polyphase
 representation of (a).}
\label{fig:interpolation}
\end{figure}

The interpolation of a decimated signal is completed 
by filtering $v=(\us{\M})(\ds{\M})x$
by a digital filter $\K$ (see Fig.\ \ref{fig:interpolation} (a)).
The decimation and interpolation process $\K(\us{\M})(\ds{\M})$ is
periodically time-varying.
To convert this equivalently to a linear time-invariant system,
we introduce the polyphase decomposition \cite{Vai}.
%more precisely, it is  $(N,M)$-periodic \cite{Mey90} or 
%$(N,M)$-shift-invariant \cite{CheQiuBai98}.
%By this fact,
%the multirate system can be equivalently represented as
%a time-invariant system 
%via discrete-time lifting \cite{CheFra},
%which is also called polyphase decomposition \cite{Vai}.
%via polyphase decomposition \cite{Vai}.
Let $\dlift{M}$ be the polyphase decomposition operator, that is,
\[
\dlift{M}: \left\{x_0,x_1,\ldots\right\}\mapsto \left\{
	\left[\begin{array}{c}x_0\\x_1\\\vdots\\x_{M-1}\end{array}\right],
	\left[\begin{array}{c}x_M\\x_{M+1}\\\vdots\\x_{2M-1}\end{array}\right],
	\ldots\right\}.
\]
By this operator, the nonuniform decimator $\ds{\M}$ and expander $\us{\M}$ can be represented
as filterbanks.
\begin{lemma}
\label{lem:polyphase}
The following two equalities hold:
\begin{enumerate}
\item $\ds{\M} = \idlift{N}E\dlift{M},$
\item $\us{\M} = \idlift{M}E^\top\dlift{N},$
\end{enumerate}
where
$E=[E_{ij}]$ is an $N\times M$ matrix whose elements are
defined as follows:
%$E_{ij} = 1$ if $(i,j)=(1,i_1), (2, i_2), \ldots, (N,i_N)$,
%otherwise $E_{ij} = 0$.
\[
E_{ij} = \begin{cases}1, \quad (i,j)=(1,i_1+1), (2, i_2+1), \ldots, (N,i_N+1),\\
		0, \quad \text{otherwise}.
	\end{cases}
\]
\end{lemma}
%\begin{proof}
\subsubsection*{Proof.}
\begin{enumerate}
\item
For $v=[v_0,v_1,\ldots,v_{M-1}]^\top\in\real^M$ we have
\[
 Ev = [v_{i_1}, v_{i_2},\ldots, v_{i_{N}}]^\top \in\real^N.
\]
By using this, for any sequence $x=\{x_0,x_1,x_2,\ldots,\}$ we have
\[
 \begin{split}
  E\dlift{M}x 
  &= E\left\{\begin{bmatrix}x_0\\x_1\\\vdots\\x_{M-1}\end{bmatrix},~ \begin{bmatrix}x_M\\x_{M+1}\\\vdots\\x_{2M-1}\end{bmatrix}, \ldots \right\}\\
  &= \left\{E\begin{bmatrix}x_0\\x_1\\\vdots\\x_{M-1}\end{bmatrix},~ E\begin{bmatrix}x_M\\x_{M+1}\\\vdots\\x_{2M-1}\end{bmatrix}, \ldots \right\}\\
  &= \left\{\begin{bmatrix}x_{i_1}\\x_{i_2}\\\vdots\\x_{i_N}\end{bmatrix},~ \begin{bmatrix}x_{M+i_1}\\x_{M+i_2}\\\vdots\\x_{M+i_N}\end{bmatrix}, \ldots \right\}.
 \end{split}
\]
Therefore, for any $x$ we have
\[
 \begin{split}
 \idlift{N}E\dlift{M}x &= \left\{x_{i_1},x_{i_2},\ldots,x_{i_N},x_{M+i_1},x_{M+i_2},\ldots,x_{M+i_N},\ldots\right\}\\
   &= (\ds{\M})x.
 \end{split}
\]
That is, $\idlift{N}E\dlift{M}=\ds{\M}$.
\item
The matrix $E$ can be represented by
\[
 E^\top = [e_{i_1},e_{i_2},\ldots,e_{i_N}],
\]
where
\[
 e_i=[0,\ldots,0, \stackrel{i+1}{\stackrel{\vee}{1}},0,\ldots,0]^\top \in \real^M, i=0,1,\ldots,M-1.
\]
By using this, for any vector of
\[
w=[w_0,w_1,\ldots,w_{N-1}]^\top \in \real^N,
\]
we have
\[
\begin{split}
E^\top w 
&= [e_{i_1},e_{i_2},\ldots,e_{i_N}] \begin{bmatrix}w_0\\w_1\\\vdots\\w_{N-1}\end{bmatrix}\\
&=e_{i_1}w_0+e_{i_2}w_1+\cdots+e_{i_N}w_{N-1}\\
&= [0,\ldots,0,\stackrel{i_1+1}{\stackrel{\vee}{w_0}},0,\ldots,0,\stackrel{i_2+1}{\stackrel{\vee}{w_1}},0,\ldots, 0, \stackrel{i_N+1}{\stackrel{\vee}{w_{N-1}}},0,\ldots,0]^\top \in\real^M.
\end{split}
\]
%%% R2-5
By this property and the same computation as in the proof of $\ds{\M}$,
the equality $\idlift{M}E^\top\dlift{N}=\us{\M}$ can be proved.
\end{enumerate}
%\end{proof}
\hfill$\Box$

\begin{figure}[tb]
\begin{center}
%%%%\input{FB_s.tex}
%WinTpicVersion3.08
\unitlength 0.1in
\begin{picture}( 30.8000, 25.8500)(  2.0000,-26.9500)
% BOX 2 0 3 0
% 2 1160 200 1480 520
% 
\special{pn 8}%
\special{pa 1160 200}%
\special{pa 1480 200}%
\special{pa 1480 520}%
\special{pa 1160 520}%
\special{pa 1160 200}%
\special{fp}%
% BOX 2 0 3 0
% 2 1940 200 2260 520
% 
\special{pn 8}%
\special{pa 1940 200}%
\special{pa 2260 200}%
\special{pa 2260 520}%
\special{pa 1940 520}%
\special{pa 1940 200}%
\special{fp}%
% BOX 2 0 3 0
% 2 1160 680 1480 1000
% 
\special{pn 8}%
\special{pa 1160 680}%
\special{pa 1480 680}%
\special{pa 1480 1000}%
\special{pa 1160 1000}%
\special{pa 1160 680}%
\special{fp}%
% BOX 2 0 3 0
% 2 1940 680 2260 1000
% 
\special{pn 8}%
\special{pa 1940 680}%
\special{pa 2260 680}%
\special{pa 2260 1000}%
\special{pa 1940 1000}%
\special{pa 1940 680}%
\special{fp}%
% VECTOR 2 0 3 0
% 2 2260 840 2420 840
% 
\special{pn 8}%
\special{pa 2260 840}%
\special{pa 2420 840}%
\special{fp}%
\special{sh 1}%
\special{pa 2420 840}%
\special{pa 2354 820}%
\special{pa 2368 840}%
\special{pa 2354 860}%
\special{pa 2420 840}%
\special{fp}%
% VECTOR 2 0 3 0
% 2 1000 840 1160 840
% 
\special{pn 8}%
\special{pa 1000 840}%
\special{pa 1160 840}%
\special{fp}%
\special{sh 1}%
\special{pa 1160 840}%
\special{pa 1094 820}%
\special{pa 1108 840}%
\special{pa 1094 860}%
\special{pa 1160 840}%
\special{fp}%
% BOX 2 0 3 0
% 2 1000 680 680 1000
% 
\special{pn 8}%
\special{pa 1000 680}%
\special{pa 680 680}%
\special{pa 680 1000}%
\special{pa 1000 1000}%
\special{pa 1000 680}%
\special{fp}%
% VECTOR 2 0 3 0
% 2 520 840 680 840
% 
\special{pn 8}%
\special{pa 520 840}%
\special{pa 680 840}%
\special{fp}%
\special{sh 1}%
\special{pa 680 840}%
\special{pa 614 820}%
\special{pa 628 840}%
\special{pa 614 860}%
\special{pa 680 840}%
\special{fp}%
% LINE 2 0 3 0
% 2 520 840 520 360
% 
\special{pn 8}%
\special{pa 520 840}%
\special{pa 520 360}%
\special{fp}%
% VECTOR 2 0 3 0
% 2 200 360 1160 360
% 
\special{pn 8}%
\special{pa 200 360}%
\special{pa 1160 360}%
\special{fp}%
\special{sh 1}%
\special{pa 1160 360}%
\special{pa 1094 340}%
\special{pa 1108 360}%
\special{pa 1094 380}%
\special{pa 1160 360}%
\special{fp}%
% VECTOR 2 0 3 0
% 2 2920 840 2920 400
% 
\special{pn 8}%
\special{pa 2920 840}%
\special{pa 2920 400}%
\special{fp}%
\special{sh 1}%
\special{pa 2920 400}%
\special{pa 2900 468}%
\special{pa 2920 454}%
\special{pa 2940 468}%
\special{pa 2920 400}%
\special{fp}%
% CIRCLE 2 0 3 0
% 4 2920 360 2920 400 2920 400 2920 400
% 
\special{pn 8}%
\special{ar 2920 360 40 40  0.0000000 6.2831853}%
% VECTOR 2 0 3 0
% 2 2960 360 3280 360
% 
\special{pn 8}%
\special{pa 2960 360}%
\special{pa 3280 360}%
\special{fp}%
\special{sh 1}%
\special{pa 3280 360}%
\special{pa 3214 340}%
\special{pa 3228 360}%
\special{pa 3214 380}%
\special{pa 3280 360}%
\special{fp}%
% STR 2 0 3 0
% 3 2880 320 2880 400 4 0
% $+$
\put(28.8000,-4.0000){\makebox(0,0)[rt]{$+$}}%
% STR 2 0 3 0
% 3 1320 280 1320 360 5 0
% $\ds{3}$
\put(13.2000,-3.6000){\makebox(0,0){$\ds{3}$}}%
% STR 2 0 3 0
% 3 2100 280 2100 360 5 0
% $\us{2}$
\put(21.0000,-3.6000){\makebox(0,0){$\us{2}$}}%
% STR 2 0 3 0
% 3 2100 760 2100 840 5 0
% $\us{2}$
\put(21.0000,-8.4000){\makebox(0,0){$\us{2}$}}%
% STR 2 0 3 0
% 3 1320 760 1320 840 5 0
% $\ds{3}$
\put(13.2000,-8.4000){\makebox(0,0){$\ds{3}$}}%
% STR 2 0 3 0
% 3 840 760 840 840 5 0
% $z$
\put(8.4000,-8.4000){\makebox(0,0){$z$}}%
% BOX 2 0 3 0
% 2 2740 680 2420 1000
% 
\special{pn 8}%
\special{pa 2740 680}%
\special{pa 2420 680}%
\special{pa 2420 1000}%
\special{pa 2740 1000}%
\special{pa 2740 680}%
\special{fp}%
% STR 2 0 3 0
% 3 2580 760 2580 840 5 0
% $z^{-1}$
\put(25.8000,-8.4000){\makebox(0,0){$z^{-1}$}}%
% VECTOR 2 0 3 0
% 2 1480 360 1940 360
% 
\special{pn 8}%
\special{pa 1480 360}%
\special{pa 1940 360}%
\special{fp}%
\special{sh 1}%
\special{pa 1940 360}%
\special{pa 1874 340}%
\special{pa 1888 360}%
\special{pa 1874 380}%
\special{pa 1940 360}%
\special{fp}%
% VECTOR 2 0 3 0
% 2 2260 360 2880 360
% 
\special{pn 8}%
\special{pa 2260 360}%
\special{pa 2880 360}%
\special{fp}%
\special{sh 1}%
\special{pa 2880 360}%
\special{pa 2814 340}%
\special{pa 2828 360}%
\special{pa 2814 380}%
\special{pa 2880 360}%
\special{fp}%
% VECTOR 2 0 3 0
% 2 1480 840 1940 840
% 
\special{pn 8}%
\special{pa 1480 840}%
\special{pa 1940 840}%
\special{fp}%
\special{sh 1}%
\special{pa 1940 840}%
\special{pa 1874 820}%
\special{pa 1888 840}%
\special{pa 1874 860}%
\special{pa 1940 840}%
\special{fp}%
% LINE 2 0 3 0
% 2 2740 840 2920 840
% 
\special{pn 8}%
\special{pa 2740 840}%
\special{pa 2920 840}%
\special{fp}%
% BOX 2 0 3 0
% 2 1160 1600 1480 1920
% 
\special{pn 8}%
\special{pa 1160 1600}%
\special{pa 1480 1600}%
\special{pa 1480 1920}%
\special{pa 1160 1920}%
\special{pa 1160 1600}%
\special{fp}%
% BOX 2 0 3 0
% 2 1940 1600 2260 1920
% 
\special{pn 8}%
\special{pa 1940 1600}%
\special{pa 2260 1600}%
\special{pa 2260 1920}%
\special{pa 1940 1920}%
\special{pa 1940 1600}%
\special{fp}%
% BOX 2 0 3 0
% 2 1160 2080 1480 2400
% 
\special{pn 8}%
\special{pa 1160 2080}%
\special{pa 1480 2080}%
\special{pa 1480 2400}%
\special{pa 1160 2400}%
\special{pa 1160 2080}%
\special{fp}%
% BOX 2 0 3 0
% 2 1940 2080 2260 2400
% 
\special{pn 8}%
\special{pa 1940 2080}%
\special{pa 2260 2080}%
\special{pa 2260 2400}%
\special{pa 1940 2400}%
\special{pa 1940 2080}%
\special{fp}%
% VECTOR 2 0 3 0
% 2 2260 2240 2420 2240
% 
\special{pn 8}%
\special{pa 2260 2240}%
\special{pa 2420 2240}%
\special{fp}%
\special{sh 1}%
\special{pa 2420 2240}%
\special{pa 2354 2220}%
\special{pa 2368 2240}%
\special{pa 2354 2260}%
\special{pa 2420 2240}%
\special{fp}%
% VECTOR 2 0 3 0
% 2 1000 2240 1160 2240
% 
\special{pn 8}%
\special{pa 1000 2240}%
\special{pa 1160 2240}%
\special{fp}%
\special{sh 1}%
\special{pa 1160 2240}%
\special{pa 1094 2220}%
\special{pa 1108 2240}%
\special{pa 1094 2260}%
\special{pa 1160 2240}%
\special{fp}%
% BOX 2 0 3 0
% 2 1000 2080 680 2400
% 
\special{pn 8}%
\special{pa 1000 2080}%
\special{pa 680 2080}%
\special{pa 680 2400}%
\special{pa 1000 2400}%
\special{pa 1000 2080}%
\special{fp}%
% VECTOR 2 0 3 0
% 2 520 2240 680 2240
% 
\special{pn 8}%
\special{pa 520 2240}%
\special{pa 680 2240}%
\special{fp}%
\special{sh 1}%
\special{pa 680 2240}%
\special{pa 614 2220}%
\special{pa 628 2240}%
\special{pa 614 2260}%
\special{pa 680 2240}%
\special{fp}%
% LINE 2 0 3 0
% 2 520 2240 520 1760
% 
\special{pn 8}%
\special{pa 520 2240}%
\special{pa 520 1760}%
\special{fp}%
% VECTOR 2 0 3 0
% 2 200 1760 1160 1760
% 
\special{pn 8}%
\special{pa 200 1760}%
\special{pa 1160 1760}%
\special{fp}%
\special{sh 1}%
\special{pa 1160 1760}%
\special{pa 1094 1740}%
\special{pa 1108 1760}%
\special{pa 1094 1780}%
\special{pa 1160 1760}%
\special{fp}%
% VECTOR 2 0 3 0
% 2 2920 2240 2920 1800
% 
\special{pn 8}%
\special{pa 2920 2240}%
\special{pa 2920 1800}%
\special{fp}%
\special{sh 1}%
\special{pa 2920 1800}%
\special{pa 2900 1868}%
\special{pa 2920 1854}%
\special{pa 2940 1868}%
\special{pa 2920 1800}%
\special{fp}%
% CIRCLE 2 0 3 0
% 4 2920 1760 2920 1800 2920 1800 2920 1800
% 
\special{pn 8}%
\special{ar 2920 1760 40 40  0.0000000 6.2831853}%
% VECTOR 2 0 3 0
% 2 2960 1760 3280 1760
% 
\special{pn 8}%
\special{pa 2960 1760}%
\special{pa 3280 1760}%
\special{fp}%
\special{sh 1}%
\special{pa 3280 1760}%
\special{pa 3214 1740}%
\special{pa 3228 1760}%
\special{pa 3214 1780}%
\special{pa 3280 1760}%
\special{fp}%
% STR 2 0 3 0
% 3 2880 1720 2880 1800 4 0
% $+$
\put(28.8000,-18.0000){\makebox(0,0)[rt]{$+$}}%
% STR 2 0 3 0
% 3 1320 1680 1320 1760 5 0
% $\ds{2}$
\put(13.2000,-17.6000){\makebox(0,0){$\ds{2}$}}%
% STR 2 0 3 0
% 3 2100 1680 2100 1760 5 0
% $\us{3}$
\put(21.0000,-17.6000){\makebox(0,0){$\us{3}$}}%
% STR 2 0 3 0
% 3 2100 2160 2100 2240 5 0
% $\us{3}$
\put(21.0000,-22.4000){\makebox(0,0){$\us{3}$}}%
% STR 2 0 3 0
% 3 1320 2160 1320 2240 5 0
% $\ds{2}$
\put(13.2000,-22.4000){\makebox(0,0){$\ds{2}$}}%
% STR 2 0 3 0
% 3 840 2160 840 2240 5 0
% $z$
\put(8.4000,-22.4000){\makebox(0,0){$z$}}%
% BOX 2 0 3 0
% 2 2740 2080 2420 2400
% 
\special{pn 8}%
\special{pa 2740 2080}%
\special{pa 2420 2080}%
\special{pa 2420 2400}%
\special{pa 2740 2400}%
\special{pa 2740 2080}%
\special{fp}%
% STR 2 0 3 0
% 3 2580 2160 2580 2240 5 0
% $z^{-1}$
\put(25.8000,-22.4000){\makebox(0,0){$z^{-1}$}}%
% VECTOR 2 0 3 0
% 2 1480 1760 1940 1760
% 
\special{pn 8}%
\special{pa 1480 1760}%
\special{pa 1940 1760}%
\special{fp}%
\special{sh 1}%
\special{pa 1940 1760}%
\special{pa 1874 1740}%
\special{pa 1888 1760}%
\special{pa 1874 1780}%
\special{pa 1940 1760}%
\special{fp}%
% VECTOR 2 0 3 0
% 2 2260 1760 2880 1760
% 
\special{pn 8}%
\special{pa 2260 1760}%
\special{pa 2880 1760}%
\special{fp}%
\special{sh 1}%
\special{pa 2880 1760}%
\special{pa 2814 1740}%
\special{pa 2828 1760}%
\special{pa 2814 1780}%
\special{pa 2880 1760}%
\special{fp}%
% VECTOR 2 0 3 0
% 2 1480 2240 1940 2240
% 
\special{pn 8}%
\special{pa 1480 2240}%
\special{pa 1940 2240}%
\special{fp}%
\special{sh 1}%
\special{pa 1940 2240}%
\special{pa 1874 2220}%
\special{pa 1888 2240}%
\special{pa 1874 2260}%
\special{pa 1940 2240}%
\special{fp}%
% LINE 2 0 3 0
% 2 2740 2240 2920 2240
% 
\special{pn 8}%
\special{pa 2740 2240}%
\special{pa 2920 2240}%
\special{fp}%
% STR 2 0 3 0
% 3 1700 1280 1700 1380 5 0
% (a)
\put(17.0000,-13.8000){\makebox(0,0){(a)}}%
% STR 2 0 3 0
% 3 1700 2680 1700 2780 5 0
% (b)
\put(17.0000,-27.8000){\makebox(0,0){(b)}}%
% STR 2 0 3 0
% 3 1680 2660 1680 2760 5 0
%  
\put(16.8000,-27.6000){\makebox(0,0){ }}%
% BOX 2 2 3 0
% 2 410 110 1610 1110
% 
\special{pn 8}%
\special{pa 410 110}%
\special{pa 1610 110}%
\special{pa 1610 1110}%
\special{pa 410 1110}%
\special{pa 410 110}%
\special{dt 0.045}%
% STR 2 0 3 0
% 3 1010 1110 1010 1210 5 0
% $E\dlift{3}$
\put(10.1000,-12.1000){\makebox(0,0){$E\dlift{3}$}}%
% BOX 2 2 3 0
% 2 1810 1510 3010 2510
% 
\special{pn 8}%
\special{pa 1810 1510}%
\special{pa 3010 1510}%
\special{pa 3010 2510}%
\special{pa 1810 2510}%
\special{pa 1810 1510}%
\special{dt 0.045}%
% STR 2 0 3 0
% 3 2410 2510 2410 2610 5 0
% $\idlift{3}E^\top$
\put(24.1000,-26.1000){\makebox(0,0){$\idlift{3}E^\top$}}%
% BOX 2 2 3 0
% 2 1810 110 3010 1110
% 
\special{pn 8}%
\special{pa 1810 110}%
\special{pa 3010 110}%
\special{pa 3010 1110}%
\special{pa 1810 1110}%
\special{pa 1810 110}%
\special{dt 0.045}%
% STR 2 0 3 0
% 3 2410 1110 2410 1210 5 0
% $\idlift{2}$
\put(24.1000,-12.1000){\makebox(0,0){$\idlift{2}$}}%
% BOX 2 2 3 0
% 2 410 1510 1610 2510
% 
\special{pn 8}%
\special{pa 410 1510}%
\special{pa 1610 1510}%
\special{pa 1610 2510}%
\special{pa 410 2510}%
\special{pa 410 1510}%
\special{dt 0.045}%
% STR 2 0 3 0
% 3 1010 2510 1010 2610 5 0
% $\dlift{2}$
\put(10.1000,-26.1000){\makebox(0,0){$\dlift{2}$}}%
\end{picture}%

\vspace{5mm}
\end{center}
\caption{Filterbank representation of (a) $\ds{\M}$ and (b) $\us{\M}$
with $\M=[1,1,0]$ ($M=3$, $N=2$).}
\label{fig:FB}
\end{figure}
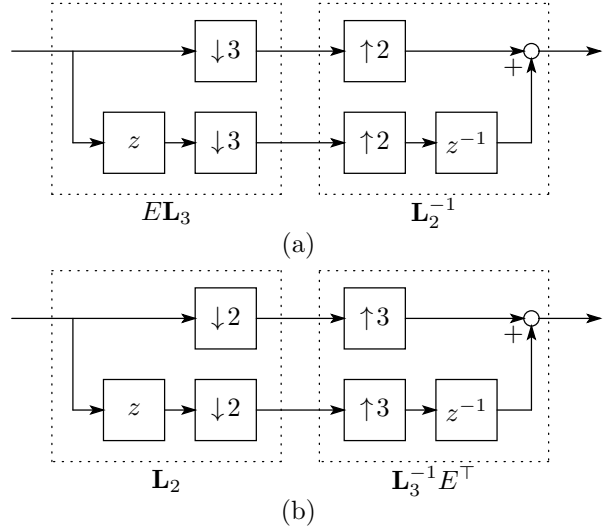

For example, if $\M=[1,1,0]$ ($M=3$, $N=2$, $i_1=1$, $i_2=2$) then 
$\us{\M}$ and $\ds{\M}$ are represented as filterbanks
shown in Fig.\ref{fig:FB}.
In this case, the matrix $E$ is given by
\[
E = \left[\begin{array}{ccc}1&0&0\\0&1&0\end{array}\right].
\]

By Lemma \ref{lem:polyphase}, we can represent the decimation and interpolation process
$\K(\us{\M})(\ds{\M})$ by a polyphase decomposition.
In fact, we have the following theorem (see also Fig.\ \ref{fig:interpolation}).
%$
%\K(\us{\M})(\ds{\M}) = \idlift{N}\tK E \dlift{M}
%$
%where
%$\tK := \dlift{N}\K\idlift{N}$,
\begin{theorem}
\label{thm:polyphase}
The following identity holds:
\begin{equation}
\K(\us{\M})(\ds{\M}) = \idlift{M}\tK E \dlift{M},\label{eq:dliftK}
\end{equation}
where
\begin{equation}
\tK := \dlift{M}\K\idlift{M}E^\top.\label{eq:tK}
\end{equation}
%and
%$D_{\M} := E^\top E = \diag\left\{b_0,b_1,\ldots,b_{M-1}\right\}$.
\end{theorem}
%\begin{proof}
\subsubsection*{Proof.}
By Lemma \ref{lem:polyphase} and the identities $\idlift{M}\dlift{M}=1$
and $\dlift{N}\idlift{N}=I_N$ ($I_N$ is the $N\times N$ identity matrix), we have
\[
 \begin{split}
  \K(\us{\M})(\ds{\M}) &= \idlift{M}\dlift{M}\K\left(\idlift{M}E^\top\dlift{N}\right)\left(\idlift{N}E\dlift{M}\right)\\
   &= \idlift{M} \left(\dlift{M}\K\idlift{M}E^\top\right)  E \dlift{M}\\
   &= \idlift{M} \tK E \dlift{M}.
 \end{split}
\]
%\end{proof}
\hfill $\Box$

By this theorem,  we can easily see that $\tK$ is a linear time-invariant system with $M$ inputs and $M$ outputs
\cite[Chap. 8]{CheFra}, and
$\K(\us{\M})(\ds{\M})$ is $M$-periodic \cite{Mey90}.
%%% R2-6
%\begin{cor}
%\begin{enumerate}
%\item
%The filter $\tK$ is a linear and time-invariant system with $M$ inputs and $M$ outputs.
%\item
%The process $\K(\us{\M})(\ds{\M})$ is $M$-periodic.
%\end{enumerate}
%\end{cor}

\subsection{$H^\infty$ Optimal Interpolation for Non-Band-Limited Signals}
We now consider the signal space to which the original continuous-time
signals before sampling and decimation belong.  
%%% R1-2, R2-8
Let $h$ denote the sampling period.
The nonuniform sampling theory \cite{VaiLiu90,Vai} assumes this space as
the band-limited subspace defined by
%%% R2-7
\[
  \BL := \left\{u \in L^2: \supp~\hat{u} \subset \Omega(M/N)\right\},
\]
where $\hat{u}$ is the Fourier transform of $u\in L^2$, and
\[
  \Omega(M/N):=\left(-\frac{N\pi}{Mh}, \frac{N\pi}{Mh}\right).
\]
On the other hand, we consider another subspace of $L^2$
which includes non-band-limited signals, defined by
\[
 \FL:=\left\{u \in L^2: u=Fw, ~ w\in L^2\right\},
\]
where $F$ is a stable linear time-invariant continuous-time system
whose transfer function is finite-dimensional and
strictly proper.
This space is a model for the signal subspace
to which the input analog signals belong.
A merit for this model is that one can naturally and easily
include the analog frequency characteristic in the model
via physical laws or executing Fourier transform of real signals.
This is an advantage over the generalized sampling theory
\cite{UnsZer98,Uns05}, in which the signal subspace is modeled
by the linear span of a given generating function.

Moreover, this subspace $\FL$ is essentially wider than $\BL$.
In fact, the following lemma holds:
\begin{lemma}
Assume that $F(\jj\omega)$ has no zeros in $\Omega(M/N)$.
Then $\BL \subset \FL$.
\end{lemma}
%\begin{proof}
\subsubsection*{Proof.}
Let $u\in \BL$. Define a function $w$ such that
\[
 \hat{w}(\jj \omega) = \begin{cases} 
		\hat{F}(\jj \omega)^{-1}\hat{u}(\jj\omega),&\quad \text{if}~ \omega \in \Omega(M/N),\\
		0, &\quad \text{if}~ \omega \notin \Omega(M/N).
 \end{cases}
\]
%%% R1-2
Since $u\in\BL$, $\hat{u}(\jj\omega)=0$ if $\omega \notin \Omega(M/N)$,
and hence we have $\hat{u}(\jj\omega)=\hat{F}(\jj\omega)\hat{w}(\jj\omega)$
for all $\omega\in\real$, or $u=Fw$.
Then we show that $w\in L^2$.
In fact, we have
\[
 \begin{split}
%\begin{Meqnarray}
  \|w\|_2^2 
  &= \int_0^\infty |w(t)|^2\dd t\\
  &=\frac{1}{2\pi}\int_{\real} |\hat{w}(\jj\omega)|^2\dd \omega\\
  &=\frac{1}{2\pi}\int_{\Omega(M/N)} |\hat{F}(\jj\omega)^{-1}u(\jj\omega)|^2\dd \omega\\
  &\leq \frac{1}{2\pi}~\|u\|_2^2~\left(\max_{\omega\in\Omega(M/N)} \left|\hat{F}(\jj\omega)^{-1}\right|\right)^2\\
  &= \frac{1}{2\pi}~\|u\|_2^2~\left(\min_{\omega\in\Omega(M/N)} \left|\hat{F}(\jj\omega)\right|^2\right)^{-1}\\
  &<\infty.
 \end{split}
\]
%\end{Meqnarray}
Therefore, we have $u\in\FL$.
%\end{proof}
\hfill $\Box$

\begin{figure}[tb]
\begin{center}
%%%%\input{errorsys.tex}
%WinTpicVersion3.08
\unitlength 0.1in
\begin{picture}( 30.3000,  6.0000)(  5.0000, -8.0000)
% BOX 2 0 3 0
% 2 680 560 920 800
% 
\special{pn 8}%
\special{pa 680 560}%
\special{pa 920 560}%
\special{pa 920 800}%
\special{pa 680 800}%
\special{pa 680 560}%
\special{fp}%
% VECTOR 2 0 3 0
% 2 920 680 1160 680
% 
\special{pn 8}%
\special{pa 920 680}%
\special{pa 1160 680}%
\special{fp}%
\special{sh 1}%
\special{pa 1160 680}%
\special{pa 1094 660}%
\special{pa 1108 680}%
\special{pa 1094 700}%
\special{pa 1160 680}%
\special{fp}%
% BOX 2 0 3 0
% 2 1160 560 1400 800
% 
\special{pn 8}%
\special{pa 1160 560}%
\special{pa 1400 560}%
\special{pa 1400 800}%
\special{pa 1160 800}%
\special{pa 1160 560}%
\special{fp}%
% VECTOR 2 0 3 0
% 2 1400 680 1520 680
% 
\special{pn 8}%
\special{pa 1400 680}%
\special{pa 1520 680}%
\special{fp}%
\special{sh 1}%
\special{pa 1520 680}%
\special{pa 1454 660}%
\special{pa 1468 680}%
\special{pa 1454 700}%
\special{pa 1520 680}%
\special{fp}%
% BOX 2 0 3 0
% 2 1520 560 1760 800
% 
\special{pn 8}%
\special{pa 1520 560}%
\special{pa 1760 560}%
\special{pa 1760 800}%
\special{pa 1520 800}%
\special{pa 1520 560}%
\special{fp}%
% VECTOR 2 0 3 0
% 2 1760 680 1880 680
% 
\special{pn 8}%
\special{pa 1760 680}%
\special{pa 1880 680}%
\special{fp}%
\special{sh 1}%
\special{pa 1880 680}%
\special{pa 1814 660}%
\special{pa 1828 680}%
\special{pa 1814 700}%
\special{pa 1880 680}%
\special{fp}%
% BOX 2 0 3 0
% 2 1880 560 2120 800
% 
\special{pn 8}%
\special{pa 1880 560}%
\special{pa 2120 560}%
\special{pa 2120 800}%
\special{pa 1880 800}%
\special{pa 1880 560}%
\special{fp}%
% VECTOR 2 0 3 0
% 2 2120 680 2240 680
% 
\special{pn 8}%
\special{pa 2120 680}%
\special{pa 2240 680}%
\special{fp}%
\special{sh 1}%
\special{pa 2240 680}%
\special{pa 2174 660}%
\special{pa 2188 680}%
\special{pa 2174 700}%
\special{pa 2240 680}%
\special{fp}%
% BOX 2 0 3 0
% 2 2240 560 2480 800
% 
\special{pn 8}%
\special{pa 2240 560}%
\special{pa 2480 560}%
\special{pa 2480 800}%
\special{pa 2240 800}%
\special{pa 2240 560}%
\special{fp}%
% VECTOR 2 0 3 0
% 2 2480 680 2600 680
% 
\special{pn 8}%
\special{pa 2480 680}%
\special{pa 2600 680}%
\special{fp}%
\special{sh 1}%
\special{pa 2600 680}%
\special{pa 2534 660}%
\special{pa 2548 680}%
\special{pa 2534 700}%
\special{pa 2600 680}%
\special{fp}%
% BOX 2 0 3 0
% 2 2600 560 2840 800
% 
\special{pn 8}%
\special{pa 2600 560}%
\special{pa 2840 560}%
\special{pa 2840 800}%
\special{pa 2600 800}%
\special{pa 2600 560}%
\special{fp}%
% VECTOR 2 0 3 0
% 2 2840 680 2960 680
% 
\special{pn 8}%
\special{pa 2840 680}%
\special{pa 2960 680}%
\special{fp}%
\special{sh 1}%
\special{pa 2960 680}%
\special{pa 2894 660}%
\special{pa 2908 680}%
\special{pa 2894 700}%
\special{pa 2960 680}%
\special{fp}%
% BOX 2 0 3 0
% 2 2960 560 3200 800
% 
\special{pn 8}%
\special{pa 2960 560}%
\special{pa 3200 560}%
\special{pa 3200 800}%
\special{pa 2960 800}%
\special{pa 2960 560}%
\special{fp}%
% VECTOR 2 0 3 0
% 2 3200 680 3320 680
% 
\special{pn 8}%
\special{pa 3200 680}%
\special{pa 3320 680}%
\special{fp}%
\special{sh 1}%
\special{pa 3320 680}%
\special{pa 3254 660}%
\special{pa 3268 680}%
\special{pa 3254 700}%
\special{pa 3320 680}%
\special{fp}%
% LINE 2 0 3 0
% 2 1040 680 1040 320
% 
\special{pn 8}%
\special{pa 1040 680}%
\special{pa 1040 320}%
\special{fp}%
% VECTOR 2 0 3 0
% 2 1040 320 1760 320
% 
\special{pn 8}%
\special{pa 1040 320}%
\special{pa 1760 320}%
\special{fp}%
\special{sh 1}%
\special{pa 1760 320}%
\special{pa 1694 300}%
\special{pa 1708 320}%
\special{pa 1694 340}%
\special{pa 1760 320}%
\special{fp}%
% BOX 2 0 3 0
% 2 1760 200 2120 440
% 
\special{pn 8}%
\special{pa 1760 200}%
\special{pa 2120 200}%
\special{pa 2120 440}%
\special{pa 1760 440}%
\special{pa 1760 200}%
\special{fp}%
% LINE 2 0 3 0
% 2 2120 320 3350 320
% 
\special{pn 8}%
\special{pa 2120 320}%
\special{pa 3350 320}%
\special{fp}%
% VECTOR 2 0 3 0
% 2 3350 320 3350 650
% 
\special{pn 8}%
\special{pa 3350 320}%
\special{pa 3350 650}%
\special{fp}%
\special{sh 1}%
\special{pa 3350 650}%
\special{pa 3370 584}%
\special{pa 3350 598}%
\special{pa 3330 584}%
\special{pa 3350 650}%
\special{fp}%
% CIRCLE 2 0 3 0
% 4 3350 680 3350 710 3350 710 3350 710
% 
\special{pn 8}%
\special{ar 3350 680 30 30  0.0000000 6.2831853}%
% STR 2 0 3 0
% 3 800 620 800 680 5 0
% $F$
\put(8.0000,-6.8000){\makebox(0,0){$F$}}%
% STR 2 0 3 0
% 3 1280 620 1280 680 5 0
% $\samp{h}$
\put(12.8000,-6.8000){\makebox(0,0){$\samp{h}$}}%
% STR 2 0 3 0
% 3 1640 620 1640 680 5 0
% $\dlift{M}$
\put(16.4000,-6.8000){\makebox(0,0){$\dlift{M}$}}%
% STR 2 0 3 0
% 3 2000 620 2000 680 5 0
% $E$
\put(20.0000,-6.8000){\makebox(0,0){$E$}}%
% STR 2 0 3 0
% 3 2360 620 2360 680 5 0
% $\tK$
\put(23.6000,-6.8000){\makebox(0,0){$\tK$}}%
% STR 2 0 3 0
% 3 2720 620 2720 680 5 0
% $\idlift{M}$
\put(27.2000,-6.8000){\makebox(0,0){$\idlift{M}$}}%
% STR 2 0 3 0
% 3 3080 620 3080 680 5 0
% $\hold{h}$
\put(30.8000,-6.8000){\makebox(0,0){$\hold{h}$}}%
% STR 2 0 3 0
% 3 1940 260 1940 320 5 0
% $e^{-Ls}$
\put(19.4000,-3.2000){\makebox(0,0){$e^{-Ls}$}}%
% STR 2 0 3 0
% 3 500 590 500 650 2 0
% $w$
\put(5.0000,-6.5000){\makebox(0,0)[lb]{$w$}}%
% $u$
\put(10.0000,-8.0000){\makebox(0,0)[lb]{$u$}}%
% STR 2 0 3 0
% 3 3410 590 3410 650 2 0
% $e$
\put(34.3000,-6.5000){\makebox(0,0)[lb]{$e$}}%
% VECTOR 2 0 3 0
% 2 530 680 680 680
% 
\special{pn 8}%
\special{pa 530 680}%
\special{pa 680 680}%
\special{fp}%
\special{sh 1}%
\special{pa 680 680}%
\special{pa 614 660}%
\special{pa 628 680}%
\special{pa 614 700}%
\special{pa 680 680}%
\special{fp}%
% VECTOR 2 0 3 0
% 2 3380 680 3530 680
% 
\special{pn 8}%
\special{pa 3380 680}%
\special{pa 3530 680}%
\special{fp}%
\special{sh 1}%
\special{pa 3530 680}%
\special{pa 3464 660}%
\special{pa 3478 680}%
\special{pa 3464 700}%
\special{pa 3530 680}%
\special{fp}%
% STR 2 0 3 0
% 3 3290 650 3290 710 4 0
% $-$
\put(33.4000,-7.1000){\makebox(0,0)[rt]{$-$}}%
\end{picture}%

\end{center}
\caption{Error system $T_{ew}(\tK,\M)$.}
\label{fig:errorsys}
\end{figure}
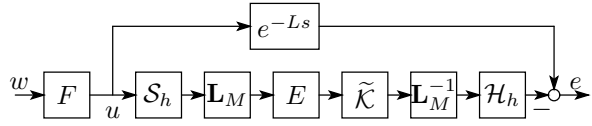

To consider signal reconstruction for non-band-limited signals in $\FL$,
let us consider the error system shown in Fig.\ \ref{fig:errorsys}.
In this figure, $F$ is a linear system
defining the signal space $\FL$.
%time-invariant continuous-time system
%whose transfer function is finite-dimensional and
%strictly proper,
%which is a model of the original analog signal.
%Since we consider reconstruction of non-band-limited signals,
%this $F$ is not the ideal lowpass filter 
%(or the Fourier transform of the sinc function).
The block $\samp{h}$ represents the ideal sampler with sampling period $h$,
and $\hold{h}$ the zero-order hold with the same sampling period.
The delay $e^{-Ls}$ is a reconstruction delay.
%\footnote{Generally, if the reconstruction delay is sufficiently large, 
%the reconstruction error can be reduced.}.
Then our optimization problem is formulated as a sampled-data $H^\infty$
optimization.
Let $T_{ew}(\tK,\M)$ be the error system from the continuous-time signal $w$
to the error $e$ (see Fig.~\ref{fig:errorsys}).
%By this notation, we emphasize that the error system depends on the
%interpolation filter $\K$ and the decimation pattern $\M$.
\begin{prob}
\label{prob:K}
Given a decimation pattern $\M$, find the optimal filter $\tK$ that minimizes
\begin{equation}
%\begin{split}
%\begin{Meqnarray}
%%% R1-3
J(\tK):=\|T_{ew}(\tK,\M)\|_\infty
 = \sup_{\begin{subarray}{c}w\in L^2\\w \neq 0\end{subarray}}
 \frac{\|T_{ew}(\tK,\M)w\|_2}{\|w\|_2}.
%\end{split}
\label{eq:hinf}
%\end{Meqnarray}
\end{equation}
\end{prob}
\subsection{Computation of Optimal Filter}
To solve Problem \ref{prob:K}, the $H^\infty$ norm
$\|T_{ew}(\tK,\M)\|_\infty$ has to be evaluated.
By using the fast discretization method, 
we can approximately obtain the optimal $\tK$ with arbitrary precision.

First we introduce useful properties \cite{CheFra} for computing the optimal filter.
\begin{lemma}
Let $\tau$ be a positive real number, and $P_c$ a continuous-time linear
time-invariant system.
Then $\ctd{\tau}{P_c}:=\samp{\tau}P_c\hold{\tau}$ is a discrete-time linear
time-invariant system.
The state space realization is given by
\begin{equation}
 \ctd{\tau}{\sspl{A}{B}{C}{D}} = \sspl{e^{A\tau}}{\int_0^\tau e^{At}B\dd t}{C}{D}.
 \label{eq:Dt}
\end{equation}
\end{lemma}
\begin{lemma}
Let $n$ be a positive integer and $P$ a discrete-time linear time-invariant system.
Then $\lift{n}{P}:=\dlift{n}P\idlift{n}$ is also a discrete-time linear time-invariant system.
The state space realization is given by
\begin{equation}
 \lift{n}{\sspl{A}{B}{C}{D}}
  = \left[\begin{array}{c|cccc}
	A^n & A^{n-1}B & A^{n-2}B & \ldots & B\\\hline
	C & D & 0 & \ldots & 0\\
	CA & CB & D & \ldots & 0\\
	\vdots & \vdots & \ddots & \vdots\\
	CA^{n-1} & CA^{n-2}B & CA^{n-3}B & \ldots & D
 \end{array}\right].
 \label{eq:Ln1}
\end{equation}
In particular, for a scalar $d\in\real$ and a matrix $D\in\real^{p\times q}$ we have respectively
%\[
% \begin{split}
%\begin{Meqnarray}
\begin{equation}
  \lift{n}{d} = d\cdot I_n,\quad
%  \lift{n}{D} &=& \mathrm{blockdiag}(D) = \begin{bmatrix}D& & \\ &\ddots& \\ & &D\end{bmatrix} \in \real^{Np\times Nq}.
  \lift{n}{D} = \mathbf{blockdiag}(\underbrace{D,\ldots,D}_{n}).
 \label{eq:Ln2}
\end{equation}
% \end{split}
%\]
%\end{Meqnarray}
\end{lemma}
Using these lemmas, we can obtain a discrete-time, 
linear and time-invariant system
whose $H^\infty$ norm approximates $\|T_{ew}(\tK,\M)\|_\infty$
with arbitrary precision. 
\begin{theorem}
\label{thm:fsfh}
Assume that $L=mh$, where $m$ is a non-negative integer.
Then 
%%% R1-5, R2-9
%for any positive integer $n$,
there exists a sequence of
%there exists a 
linear time-invariant discrete-time systems \{$T_{n}(\tK,\M)\}$ such that
\begin{equation}
 \lim_{n\rightarrow\infty} \|T_{n}(\tK,\M) \|_\infty = \|T_{ew}(\tK,\M)\|_\infty.
 \label{eq:fsfh}
\end{equation}
\end{theorem}
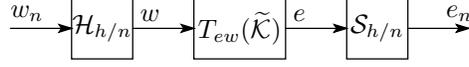
\begin{figure}[tb]
\begin{center}
%%%%\input{fsfh_s.tex}
%WinTpicVersion3.08
\unitlength 0.1in
\begin{picture}( 24.6000,  3.7000)(  3.4000, -5.1600)
% VECTOR 2 0 3 0
% 2 400 356 720 356
% 
\special{pn 8}%
\special{pa 400 356}%
\special{pa 720 356}%
\special{fp}%
\special{sh 1}%
\special{pa 720 356}%
\special{pa 654 336}%
\special{pa 668 356}%
\special{pa 654 376}%
\special{pa 720 356}%
\special{fp}%
% BOX 2 0 3 0
% 2 720 196 1040 516
% 
\special{pn 8}%
\special{pa 720 196}%
\special{pa 1040 196}%
\special{pa 1040 516}%
\special{pa 720 516}%
\special{pa 720 196}%
\special{fp}%
% VECTOR 2 0 3 0
% 2 1040 356 1360 356
% 
\special{pn 8}%
\special{pa 1040 356}%
\special{pa 1360 356}%
\special{fp}%
\special{sh 1}%
\special{pa 1360 356}%
\special{pa 1294 336}%
\special{pa 1308 356}%
\special{pa 1294 376}%
\special{pa 1360 356}%
\special{fp}%
% BOX 2 0 3 0
% 2 1360 196 1840 516
% 
\special{pn 8}%
\special{pa 1360 196}%
\special{pa 1840 196}%
\special{pa 1840 516}%
\special{pa 1360 516}%
\special{pa 1360 196}%
\special{fp}%
% VECTOR 2 0 3 0
% 2 1840 356 2160 356
% 
\special{pn 8}%
\special{pa 1840 356}%
\special{pa 2160 356}%
\special{fp}%
\special{sh 1}%
\special{pa 2160 356}%
\special{pa 2094 336}%
\special{pa 2108 356}%
\special{pa 2094 376}%
\special{pa 2160 356}%
\special{fp}%
% BOX 2 0 3 0
% 2 2160 196 2480 516
% 
\special{pn 8}%
\special{pa 2160 196}%
\special{pa 2480 196}%
\special{pa 2480 516}%
\special{pa 2160 516}%
\special{pa 2160 196}%
\special{fp}%
% VECTOR 2 0 3 0
% 2 2480 356 2800 356
% 
\special{pn 8}%
\special{pa 2480 356}%
\special{pa 2800 356}%
\special{fp}%
\special{sh 1}%
\special{pa 2800 356}%
\special{pa 2734 336}%
\special{pa 2748 356}%
\special{pa 2734 376}%
\special{pa 2800 356}%
\special{fp}%
% STR 2 0 3 0
% 3 400 236 400 316 2 0
% $w_n$
\put(4.0000,-3.1600){\makebox(0,0)[lb]{$w_n$}}%
% STR 2 0 3 0
% 3 1080 236 1080 316 2 0
% $w$
\put(10.8000,-3.1600){\makebox(0,0)[lb]{$w$}}%
% STR 2 0 3 0
% 3 1880 236 1880 316 2 0
% $e$
\put(18.8000,-3.1600){\makebox(0,0)[lb]{$e$}}%
% STR 2 0 3 0
% 3 2680 236 2680 316 2 0
% $e_n$
\put(26.8000,-3.1600){\makebox(0,0)[lb]{$e_n$}}%
% STR 2 0 3 0
% 3 2320 276 2320 356 5 0
% $\samp{h/n}$
\put(23.2000,-3.5600){\makebox(0,0){$\samp{h/n}$}}%
% STR 2 0 3 0
% 3 880 276 880 356 5 0
% $\hold{h/n}$
\put(8.8000,-3.5600){\makebox(0,0){$\hold{h/n}$}}%
% STR 2 0 3 0
% 3 1600 276 1600 356 5 0
% $T_{ew}(\tK)$
\put(16.0000,-3.5600){\makebox(0,0){$T_{ew}(\tK)$}}%
\end{picture}%

\end{center}
\caption{Fast discretization for $T_{ew}(\tK,\M)$.}
\label{fig:fsfh}
\end{figure}
\newcommand{\Gpl}{\begin{bmatrix}G_{11}&G_{12}\\G_{21}&0\end{bmatrix}}
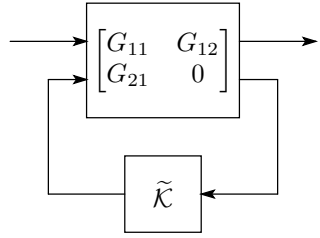
\begin{figure}[tbp]
\begin{center}
\unitlength 0.1in
\begin{picture}( 16.0000, 12.0000)(  4.0000,-14.0000)
% VECTOR 2 0 3 0
% 2 400 400 800 400
% 
\special{pn 8}%
\special{pa 400 400}%
\special{pa 800 400}%
\special{fp}%
\special{sh 1}%
\special{pa 800 400}%
\special{pa 734 380}%
\special{pa 748 400}%
\special{pa 734 420}%
\special{pa 800 400}%
\special{fp}%
% BOX 2 0 3 0
% 2 800 200 1600 800
% 
\special{pn 8}%
\special{pa 800 200}%
\special{pa 1600 200}%
\special{pa 1600 800}%
\special{pa 800 800}%
\special{pa 800 200}%
\special{fp}%
% VECTOR 2 0 3 0
% 2 1600 400 2000 400
% 
\special{pn 8}%
\special{pa 1600 400}%
\special{pa 2000 400}%
\special{fp}%
\special{sh 1}%
\special{pa 2000 400}%
\special{pa 1934 380}%
\special{pa 1948 400}%
\special{pa 1934 420}%
\special{pa 2000 400}%
\special{fp}%
% STR 2 0 3 0
% 3 1200 400 1200 500 5 0
% $\Gpl$
\put(12.0000,-5.0000){\makebox(0,0){$\Gpl$}}%
% LINE 2 0 3 0
% 4 1600 600 1800 600 1800 600 1800 1200
% 
\special{pn 8}%
\special{pa 1600 600}%
\special{pa 1800 600}%
\special{fp}%
\special{pa 1800 600}%
\special{pa 1800 1200}%
\special{fp}%
% VECTOR 2 0 3 0
% 2 1800 1200 1400 1200
% 
\special{pn 8}%
\special{pa 1800 1200}%
\special{pa 1400 1200}%
\special{fp}%
\special{sh 1}%
\special{pa 1400 1200}%
\special{pa 1468 1220}%
\special{pa 1454 1200}%
\special{pa 1468 1180}%
\special{pa 1400 1200}%
\special{fp}%
% BOX 2 0 3 0
% 2 1400 1000 1000 1400
% 
\special{pn 8}%
\special{pa 1400 1000}%
\special{pa 1000 1000}%
\special{pa 1000 1400}%
\special{pa 1400 1400}%
\special{pa 1400 1000}%
\special{fp}%
% LINE 2 0 3 0
% 4 1000 1200 600 1200 600 1200 600 600
% 
\special{pn 8}%
\special{pa 1000 1200}%
\special{pa 600 1200}%
\special{fp}%
\special{pa 600 1200}%
\special{pa 600 600}%
\special{fp}%
% VECTOR 2 0 3 0
% 2 600 600 800 600
% 
\special{pn 8}%
\special{pa 600 600}%
\special{pa 800 600}%
\special{fp}%
\special{sh 1}%
\special{pa 800 600}%
\special{pa 734 580}%
\special{pa 748 600}%
\special{pa 734 620}%
\special{pa 800 600}%
\special{fp}%
% STR 2 0 3 0
% 3 1200 1100 1200 1200 5 0
% $\tK$
\put(12.0000,-12.0000){\makebox(0,0){$\tK$}}%
\end{picture}%

\end{center}
\caption{Block diagram for discrete-time $H^\infty$ optimization.}
\label{fig:gplant}
\end{figure}
%\begin{proof}
\subsubsection*{Proof.}
We first approximate continuous-time signals $w$ and $e$ (see Fig.~\ref{fig:errorsys})
to discrete-time ones via a fast sampler $\samp{h/n}$ and a fast hold $\hold{h/n}$
(see Fig.~\ref{fig:fsfh}).
Let $G_n$ be the system from $w_n$ to $e_n$ shown in Fig.~\ref{fig:fsfh}.
Then we have
\[
 \begin{split}
 G_n &= \samp{h/n}T_{ew}(\tK,\M)\hold{h/n}\\
  &= \samp{h/n}\left\{e^{-mhs}-\hold{h}\idlift{M}\tK E\dlift{M}\samp{h}\right\}F\hold{h/n}\\
  &= z^{-mn}\samp{h/n}F\hold{h/n}-\samp{h/n}\hold{h}\idlift{M}\tK E\dlift{M}\samp{h}F\hold{h/n}.
 \end{split}
\]
%%% R2-10
Now apply the operators $\dlift{n}$ and $\idlift{n}$ to $G_n$.
Using the identities \cite[Chap.~8]{CheFra}%, we obtain 
\[
 \hold{h} = \hold{h/n}\idlift{n}H,\quad \samp{h}= S\dlift{n}\samp{h/n},
\]
where
\[
 H := [\underbrace{1,1,\ldots,1}_{n}]^\top, \quad S := [1, \underbrace{0,\ldots,0}_{n-1}],
\]
we obtain
\begin{equation}
\begin{split}
 \dlift{n}G_n\idlift{n} 
 &= \dlift{n}z^{-mn}\samp{h/n}F\hold{h/n}\idlift{n}
	- \dlift{n}\samp{h/n}\hold{h}\idlift{M}\tK E \dlift{M}\samp{h}F\hold{h/n}\idlift{n}\\
 &= G_{1,n} - G_{2,n}\idlift{M}\tK E \dlift{M}G_{3,n},\\
 G_{1,n}&:= \lift{n}{z^{-mn}\ctd{h/n}{F}},\\
% G_{2,n}&:=& \lift{n}{1}H,\\
 G_{2,n}&:= H,\\
 G_{3,n}&:= S\lift{n}{\ctd{h/n}{F}}.  \label{eq:Gn}
\end{split}
\end{equation}
Note that since $\dlift{n}$ and $\idlift{n}$ are isometric operators (with respect to $H^\infty$ norm),
the above transform preserves the norm, that is, $\|G_n\|_\infty = \|\dlift{n}G_n\idlift{n}\|_\infty$.
Then we apply $\dlift{M}$ and $\idlift{M}$ to $\dlift{n}G_n\idlift{n}$ again
to obtain 
\[
 \begin{split}
  \dlift{M}\dlift{n} G_n \idlift{n}\idlift{M} 
   &= \lift{M}{G_{1,n}} - \lift{M}{G_{2,n}}\tK E \lift{M}{G_{3,n}}\\
   &=: T_n(\tK,\M).
 \end{split}
\]
This system is a discrete-time linear time-invariant system.
The convergence property (\ref{eq:fsfh}) is shown in \cite{YamMadAnd99}.
%\end{proof}
\hfill $\Box$

%%% R2-1
The proof of Theorem \ref{thm:fsfh} gives a design procedure of the optimal filter $\tK$.
The procedure is as follows:
\begin{enumerate}
\item Compute $G_{1,n}$, $G_{2,n}$, and $G_{3,n}$
given in (\ref{eq:Gn}) by using the formulae (\ref{eq:Dt}), (\ref{eq:Ln1}), and (\ref{eq:Ln2}).
\item Compute $G_{11}:=\lift{M}{G_{1,n}}$, $G_{12}:=-\lift{M}{G_{2,n}}$,
and $G_{21}:=E\lift{M}{G_{3,n}}$ by using the formula (\ref{eq:Ln1}) and (\ref{eq:Ln2}).
%Then  the approximated system $T_n(\tK,\M)$ is given by
%\[
% T_n(\tK,\M) = \Gpl.
%\]
%
\item Solve the standard discrete-time $H^\infty$ optimal control problem
depicted in Fig.~\ref{fig:gplant} to obtain the optimal filter $\tK$.
\end{enumerate}

One can also download the MATLAB codes for obtaining the optimal filter $\tK$ through the web-page \cite{MAT}.
%%%

%%% R2-1-2, R2-1-3
Note that the fast-discretization ratio $n$ is chosen empirically.
In many cases, $n=4$ or $5$ is sufficient.
A theoretical relation between the number $n$ and the performance
is analyzed in \cite{YamAndNag02}.
Note also that the order of $\widetilde{K}$ is proportional to $n$ since the order of
the plant is proportional to $n$.
However, the filter is stable, and can be approximated by an FIR filter \cite{YamAndNagKoy03}.
%The order of this FIR filter can be chosen independently of $n$, and hence,
%the resulting filter will not be so complicated.
In many cases, the impulse response of the optimal filter decays rapidly
and the filter can be approximated almost irrespectively of $n$.
See also our example in Section 5.
%%%

%The optimization by using the fast discretization $T_n(\tK,\M)$ can be computed
%by a numerical software, e.g., MATLAB.  
%We omit the state-space formula for $T_n$ due to the limitation of space; 
%however one can download the MATLAB codes for obtaining the system $T_n$ and
%the optimal filter $\tK$ through the web-page \cite{MAT}.

\subsection{Implementation}
Once the filter $\tK$ is obtained, we can interpolate the decimated signal $(\ds{\M})x$ by
\[
\K(\us{\M})(\ds{\M})x = \left(\idlift{M}\tK\dlift{N}\right)(\ds{\M})x
\]
%Then the interpolation system is $\K(\us{\M})$, see Fig.\ref{fig:interpolation} (a) and the equation (\ref{eq:dliftK}).

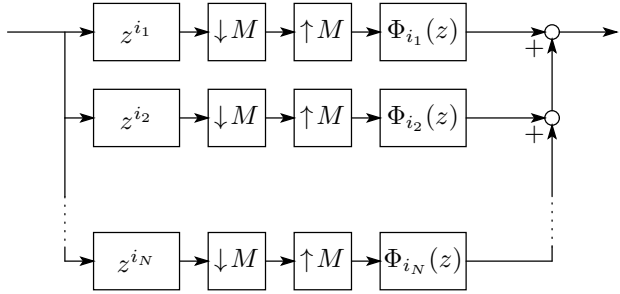
\begin{figure}[tb]
\begin{center}
%%%%\input{nonuniform_filterbank_s.tex}
%WinTpicVersion3.08
\unitlength 0.1in
\begin{picture}( 31.8800, 15.0000)(  4.0000,-17.0000)
% VECTOR 2 0 3 0
% 2 400 350 850 350
% 
\special{pn 8}%
\special{pa 400 350}%
\special{pa 850 350}%
\special{fp}%
\special{sh 1}%
\special{pa 850 350}%
\special{pa 784 330}%
\special{pa 798 350}%
\special{pa 784 370}%
\special{pa 850 350}%
\special{fp}%
% BOX 2 0 3 0
% 2 850 200 1300 500
% 
\special{pn 8}%
\special{pa 850 200}%
\special{pa 1300 200}%
\special{pa 1300 500}%
\special{pa 850 500}%
\special{pa 850 200}%
\special{fp}%
% VECTOR 2 0 3 0
% 2 1300 350 1450 350
% 
\special{pn 8}%
\special{pa 1300 350}%
\special{pa 1450 350}%
\special{fp}%
\special{sh 1}%
\special{pa 1450 350}%
\special{pa 1384 330}%
\special{pa 1398 350}%
\special{pa 1384 370}%
\special{pa 1450 350}%
\special{fp}%
% BOX 2 0 3 0
% 2 1450 200 1750 500
% 
\special{pn 8}%
\special{pa 1450 200}%
\special{pa 1750 200}%
\special{pa 1750 500}%
\special{pa 1450 500}%
\special{pa 1450 200}%
\special{fp}%
% VECTOR 2 0 3 0
% 2 1750 350 1900 350
% 
\special{pn 8}%
\special{pa 1750 350}%
\special{pa 1900 350}%
\special{fp}%
\special{sh 1}%
\special{pa 1900 350}%
\special{pa 1834 330}%
\special{pa 1848 350}%
\special{pa 1834 370}%
\special{pa 1900 350}%
\special{fp}%
% BOX 2 0 3 0
% 2 1900 200 2200 500
% 
\special{pn 8}%
\special{pa 1900 200}%
\special{pa 2200 200}%
\special{pa 2200 500}%
\special{pa 1900 500}%
\special{pa 1900 200}%
\special{fp}%
% VECTOR 2 0 3 0
% 2 2200 350 2350 350
% 
\special{pn 8}%
\special{pa 2200 350}%
\special{pa 2350 350}%
\special{fp}%
\special{sh 1}%
\special{pa 2350 350}%
\special{pa 2284 330}%
\special{pa 2298 350}%
\special{pa 2284 370}%
\special{pa 2350 350}%
\special{fp}%
% BOX 2 0 3 0
% 2 2350 200 2800 500
% 
\special{pn 8}%
\special{pa 2350 200}%
\special{pa 2800 200}%
\special{pa 2800 500}%
\special{pa 2350 500}%
\special{pa 2350 200}%
\special{fp}%
% VECTOR 2 0 3 0
% 2 2800 350 3212 350
% 
\special{pn 8}%
\special{pa 2800 350}%
\special{pa 3212 350}%
\special{fp}%
\special{sh 1}%
\special{pa 3212 350}%
\special{pa 3146 330}%
\special{pa 3160 350}%
\special{pa 3146 370}%
\special{pa 3212 350}%
\special{fp}%
% LINE 2 0 3 0
% 2 700 350 700 800
% 
\special{pn 8}%
\special{pa 700 350}%
\special{pa 700 800}%
\special{fp}%
% VECTOR 2 0 3 0
% 2 700 800 850 800
% 
\special{pn 8}%
\special{pa 700 800}%
\special{pa 850 800}%
\special{fp}%
\special{sh 1}%
\special{pa 850 800}%
\special{pa 784 780}%
\special{pa 798 800}%
\special{pa 784 820}%
\special{pa 850 800}%
\special{fp}%
% BOX 2 0 3 0
% 2 850 650 1300 950
% 
\special{pn 8}%
\special{pa 850 650}%
\special{pa 1300 650}%
\special{pa 1300 950}%
\special{pa 850 950}%
\special{pa 850 650}%
\special{fp}%
% VECTOR 2 0 3 0
% 2 1300 800 1450 800
% 
\special{pn 8}%
\special{pa 1300 800}%
\special{pa 1450 800}%
\special{fp}%
\special{sh 1}%
\special{pa 1450 800}%
\special{pa 1384 780}%
\special{pa 1398 800}%
\special{pa 1384 820}%
\special{pa 1450 800}%
\special{fp}%
% BOX 2 0 3 0
% 2 1450 650 1750 950
% 
\special{pn 8}%
\special{pa 1450 650}%
\special{pa 1750 650}%
\special{pa 1750 950}%
\special{pa 1450 950}%
\special{pa 1450 650}%
\special{fp}%
% VECTOR 2 0 3 0
% 2 1750 800 1900 800
% 
\special{pn 8}%
\special{pa 1750 800}%
\special{pa 1900 800}%
\special{fp}%
\special{sh 1}%
\special{pa 1900 800}%
\special{pa 1834 780}%
\special{pa 1848 800}%
\special{pa 1834 820}%
\special{pa 1900 800}%
\special{fp}%
% BOX 2 0 3 0
% 2 1900 650 2200 950
% 
\special{pn 8}%
\special{pa 1900 650}%
\special{pa 2200 650}%
\special{pa 2200 950}%
\special{pa 1900 950}%
\special{pa 1900 650}%
\special{fp}%
% VECTOR 2 0 3 0
% 2 2200 800 2350 800
% 
\special{pn 8}%
\special{pa 2200 800}%
\special{pa 2350 800}%
\special{fp}%
\special{sh 1}%
\special{pa 2350 800}%
\special{pa 2284 780}%
\special{pa 2298 800}%
\special{pa 2284 820}%
\special{pa 2350 800}%
\special{fp}%
% BOX 2 0 3 0
% 2 2350 650 2800 950
% 
\special{pn 8}%
\special{pa 2350 650}%
\special{pa 2800 650}%
\special{pa 2800 950}%
\special{pa 2350 950}%
\special{pa 2350 650}%
\special{fp}%
% VECTOR 2 0 3 0
% 2 2800 800 3212 800
% 
\special{pn 8}%
\special{pa 2800 800}%
\special{pa 3212 800}%
\special{fp}%
\special{sh 1}%
\special{pa 3212 800}%
\special{pa 3146 780}%
\special{pa 3160 800}%
\special{pa 3146 820}%
\special{pa 3212 800}%
\special{fp}%
% LINE 2 0 3 0
% 2 700 800 700 1212
% 
\special{pn 8}%
\special{pa 700 800}%
\special{pa 700 1212}%
\special{fp}%
% LINE 2 0 3 0
% 2 700 1475 700 1550
% 
\special{pn 8}%
\special{pa 700 1476}%
\special{pa 700 1550}%
\special{fp}%
% VECTOR 2 0 3 0
% 2 700 1550 850 1550
% 
\special{pn 8}%
\special{pa 700 1550}%
\special{pa 850 1550}%
\special{fp}%
\special{sh 1}%
\special{pa 850 1550}%
\special{pa 784 1530}%
\special{pa 798 1550}%
\special{pa 784 1570}%
\special{pa 850 1550}%
\special{fp}%
% BOX 2 0 3 0
% 2 850 1400 1300 1700
% 
\special{pn 8}%
\special{pa 850 1400}%
\special{pa 1300 1400}%
\special{pa 1300 1700}%
\special{pa 850 1700}%
\special{pa 850 1400}%
\special{fp}%
% VECTOR 2 0 3 0
% 2 1300 1550 1450 1550
% 
\special{pn 8}%
\special{pa 1300 1550}%
\special{pa 1450 1550}%
\special{fp}%
\special{sh 1}%
\special{pa 1450 1550}%
\special{pa 1384 1530}%
\special{pa 1398 1550}%
\special{pa 1384 1570}%
\special{pa 1450 1550}%
\special{fp}%
% BOX 2 0 3 0
% 2 1450 1400 1750 1700
% 
\special{pn 8}%
\special{pa 1450 1400}%
\special{pa 1750 1400}%
\special{pa 1750 1700}%
\special{pa 1450 1700}%
\special{pa 1450 1400}%
\special{fp}%
% VECTOR 2 0 3 0
% 2 1750 1550 1900 1550
% 
\special{pn 8}%
\special{pa 1750 1550}%
\special{pa 1900 1550}%
\special{fp}%
\special{sh 1}%
\special{pa 1900 1550}%
\special{pa 1834 1530}%
\special{pa 1848 1550}%
\special{pa 1834 1570}%
\special{pa 1900 1550}%
\special{fp}%
% BOX 2 0 3 0
% 2 1900 1400 2200 1700
% 
\special{pn 8}%
\special{pa 1900 1400}%
\special{pa 2200 1400}%
\special{pa 2200 1700}%
\special{pa 1900 1700}%
\special{pa 1900 1400}%
\special{fp}%
% VECTOR 2 0 3 0
% 2 2200 1550 2350 1550
% 
\special{pn 8}%
\special{pa 2200 1550}%
\special{pa 2350 1550}%
\special{fp}%
\special{sh 1}%
\special{pa 2350 1550}%
\special{pa 2284 1530}%
\special{pa 2298 1550}%
\special{pa 2284 1570}%
\special{pa 2350 1550}%
\special{fp}%
% BOX 2 0 3 0
% 2 2350 1400 2800 1700
% 
\special{pn 8}%
\special{pa 2350 1400}%
\special{pa 2800 1400}%
\special{pa 2800 1700}%
\special{pa 2350 1700}%
\special{pa 2350 1400}%
\special{fp}%
% LINE 2 0 3 0
% 2 2800 1550 3250 1550
% 
\special{pn 8}%
\special{pa 2800 1550}%
\special{pa 3250 1550}%
\special{fp}%
% CIRCLE 2 0 3 0
% 4 3250 800 3250 838 3250 838 3250 838
% 
\special{pn 8}%
\special{ar 3250 800 38 38  0.0000000 6.2831853}%
% CIRCLE 2 0 3 0
% 4 3250 350 3250 388 3250 388 3250 388
% 
\special{pn 8}%
\special{ar 3250 350 38 38  0.0000000 6.2831853}%
% VECTOR 2 0 3 0
% 2 3250 762 3250 388
% 
\special{pn 8}%
\special{pa 3250 762}%
\special{pa 3250 388}%
\special{fp}%
\special{sh 1}%
\special{pa 3250 388}%
\special{pa 3230 456}%
\special{pa 3250 442}%
\special{pa 3270 456}%
\special{pa 3250 388}%
\special{fp}%
% VECTOR 2 0 3 0
% 2 3250 1212 3250 838
% 
\special{pn 8}%
\special{pa 3250 1212}%
\special{pa 3250 838}%
\special{fp}%
\special{sh 1}%
\special{pa 3250 838}%
\special{pa 3230 906}%
\special{pa 3250 892}%
\special{pa 3270 906}%
\special{pa 3250 838}%
\special{fp}%
% LINE 2 2 3 0
% 2 700 1212 700 1475
% 
\special{pn 8}%
\special{pa 700 1212}%
\special{pa 700 1476}%
\special{dt 0.045}%
% LINE 2 0 3 0
% 2 3250 1550 3250 1475
% 
\special{pn 8}%
\special{pa 3250 1550}%
\special{pa 3250 1476}%
\special{fp}%
% LINE 2 2 3 0
% 2 3250 1475 3250 1212
% 
\special{pn 8}%
\special{pa 3250 1476}%
\special{pa 3250 1212}%
\special{dt 0.045}%
% VECTOR 2 0 3 0
% 2 3288 350 3588 350
% 
\special{pn 8}%
\special{pa 3288 350}%
\special{pa 3588 350}%
\special{fp}%
\special{sh 1}%
\special{pa 3588 350}%
\special{pa 3522 330}%
\special{pa 3536 350}%
\special{pa 3522 370}%
\special{pa 3588 350}%
\special{fp}%
% STR 2 0 3 0
% 3 1075 275 1075 350 5 0
% $z^{i_1}$
\put(10.7500,-3.5000){\makebox(0,0){$z^{i_1}$}}%
% STR 2 0 3 0
% 3 1075 725 1075 800 5 0
% $z^{i_2}$
\put(10.7500,-8.0000){\makebox(0,0){$z^{i_2}$}}%
% STR 2 0 3 0
% 3 1075 1475 1075 1550 5 0
% $z^{i_N}$
\put(10.7500,-15.5000){\makebox(0,0){$z^{i_N}$}}%
% STR 2 0 3 0
% 3 1600 275 1600 350 5 0
% $\ds{M}$
\put(16.0000,-3.5000){\makebox(0,0){$\ds{M}$}}%
% STR 2 0 3 0
% 3 1600 725 1600 800 5 0
% $\ds{M}$
\put(16.0000,-8.0000){\makebox(0,0){$\ds{M}$}}%
% STR 2 0 3 0
% 3 1600 1475 1600 1550 5 0
% $\ds{M}$
\put(16.0000,-15.5000){\makebox(0,0){$\ds{M}$}}%
% STR 2 0 3 0
% 3 2050 1475 2050 1550 5 0
% $\us{M}$
\put(20.5000,-15.5000){\makebox(0,0){$\us{M}$}}%
% STR 2 0 3 0
% 3 2050 725 2050 800 5 0
% $\us{M}$
\put(20.5000,-8.0000){\makebox(0,0){$\us{M}$}}%
% STR 2 0 3 0
% 3 2050 275 2050 350 5 0
% $\us{M}$
\put(20.5000,-3.5000){\makebox(0,0){$\us{M}$}}%
% STR 2 0 3 0
% 3 2575 275 2575 350 5 0
% $\Phi_{i_1}(z)$
\put(25.7500,-3.5000){\makebox(0,0){$\Phi_{i_1}(z)$}}%
% STR 2 0 3 0
% 3 2575 725 2575 800 5 0
% $\Phi_{i_2}(z)$
\put(25.7500,-8.0000){\makebox(0,0){$\Phi_{i_2}(z)$}}%
% STR 2 0 3 0
% 3 2575 1475 2575 1550 5 0
% $\Phi_{i_N}(z)$
\put(25.7500,-15.5000){\makebox(0,0){$\Phi_{i_N}(z)$}}%
% STR 2 0 3 0
% 3 3212 312 3212 388 4 0
% $+$
\put(32.1200,-3.8800){\makebox(0,0)[rt]{$+$}}%
% STR 2 0 3 0
% 3 3212 762 3212 838 4 0
% $+$
\put(32.1200,-8.3800){\makebox(0,0)[rt]{$+$}}%
\end{picture}%

\end{center}
\caption{Nonuniform filterbank.}
\label{fig:nonuniform-filterbank}
\end{figure}

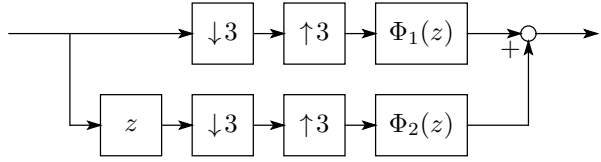
\begin{figure}[tb]
\begin{center}
%%%%\input{nonuniform_filterbank3_s.tex}
%WinTpicVersion3.08
\unitlength 0.1in
\begin{picture}( 30.8000,  8.0000)(  2.0000,-10.0000)
% BOX 2 0 3 0
% 2 1160 200 1480 520
% 
\special{pn 8}%
\special{pa 1160 200}%
\special{pa 1480 200}%
\special{pa 1480 520}%
\special{pa 1160 520}%
\special{pa 1160 200}%
\special{fp}%
% VECTOR 2 0 3 0
% 2 1480 360 1640 360
% 
\special{pn 8}%
\special{pa 1480 360}%
\special{pa 1640 360}%
\special{fp}%
\special{sh 1}%
\special{pa 1640 360}%
\special{pa 1574 340}%
\special{pa 1588 360}%
\special{pa 1574 380}%
\special{pa 1640 360}%
\special{fp}%
% BOX 2 0 3 0
% 2 1640 200 1960 520
% 
\special{pn 8}%
\special{pa 1640 200}%
\special{pa 1960 200}%
\special{pa 1960 520}%
\special{pa 1640 520}%
\special{pa 1640 200}%
\special{fp}%
% VECTOR 2 0 3 0
% 2 1960 360 2120 360
% 
\special{pn 8}%
\special{pa 1960 360}%
\special{pa 2120 360}%
\special{fp}%
\special{sh 1}%
\special{pa 2120 360}%
\special{pa 2054 340}%
\special{pa 2068 360}%
\special{pa 2054 380}%
\special{pa 2120 360}%
\special{fp}%
% BOX 2 0 3 0
% 2 2120 200 2600 520
% 
\special{pn 8}%
\special{pa 2120 200}%
\special{pa 2600 200}%
\special{pa 2600 520}%
\special{pa 2120 520}%
\special{pa 2120 200}%
\special{fp}%
% BOX 2 0 3 0
% 2 1160 680 1480 1000
% 
\special{pn 8}%
\special{pa 1160 680}%
\special{pa 1480 680}%
\special{pa 1480 1000}%
\special{pa 1160 1000}%
\special{pa 1160 680}%
\special{fp}%
% VECTOR 2 0 3 0
% 2 1480 840 1640 840
% 
\special{pn 8}%
\special{pa 1480 840}%
\special{pa 1640 840}%
\special{fp}%
\special{sh 1}%
\special{pa 1640 840}%
\special{pa 1574 820}%
\special{pa 1588 840}%
\special{pa 1574 860}%
\special{pa 1640 840}%
\special{fp}%
% BOX 2 0 3 0
% 2 1640 680 1960 1000
% 
\special{pn 8}%
\special{pa 1640 680}%
\special{pa 1960 680}%
\special{pa 1960 1000}%
\special{pa 1640 1000}%
\special{pa 1640 680}%
\special{fp}%
% VECTOR 2 0 3 0
% 2 1960 840 2120 840
% 
\special{pn 8}%
\special{pa 1960 840}%
\special{pa 2120 840}%
\special{fp}%
\special{sh 1}%
\special{pa 2120 840}%
\special{pa 2054 820}%
\special{pa 2068 840}%
\special{pa 2054 860}%
\special{pa 2120 840}%
\special{fp}%
% BOX 2 0 3 0
% 2 2120 680 2600 1000
% 
\special{pn 8}%
\special{pa 2120 680}%
\special{pa 2600 680}%
\special{pa 2600 1000}%
\special{pa 2120 1000}%
\special{pa 2120 680}%
\special{fp}%
% VECTOR 2 0 3 0
% 2 1000 840 1160 840
% 
\special{pn 8}%
\special{pa 1000 840}%
\special{pa 1160 840}%
\special{fp}%
\special{sh 1}%
\special{pa 1160 840}%
\special{pa 1094 820}%
\special{pa 1108 840}%
\special{pa 1094 860}%
\special{pa 1160 840}%
\special{fp}%
% BOX 2 0 3 0
% 2 1000 680 680 1000
% 
\special{pn 8}%
\special{pa 1000 680}%
\special{pa 680 680}%
\special{pa 680 1000}%
\special{pa 1000 1000}%
\special{pa 1000 680}%
\special{fp}%
% VECTOR 2 0 3 0
% 2 520 840 680 840
% 
\special{pn 8}%
\special{pa 520 840}%
\special{pa 680 840}%
\special{fp}%
\special{sh 1}%
\special{pa 680 840}%
\special{pa 614 820}%
\special{pa 628 840}%
\special{pa 614 860}%
\special{pa 680 840}%
\special{fp}%
% LINE 2 0 3 0
% 2 520 840 520 360
% 
\special{pn 8}%
\special{pa 520 840}%
\special{pa 520 360}%
\special{fp}%
% VECTOR 2 0 3 0
% 2 200 360 1160 360
% 
\special{pn 8}%
\special{pa 200 360}%
\special{pa 1160 360}%
\special{fp}%
\special{sh 1}%
\special{pa 1160 360}%
\special{pa 1094 340}%
\special{pa 1108 360}%
\special{pa 1094 380}%
\special{pa 1160 360}%
\special{fp}%
% VECTOR 2 0 3 0
% 2 2600 360 2880 360
% 
\special{pn 8}%
\special{pa 2600 360}%
\special{pa 2880 360}%
\special{fp}%
\special{sh 1}%
\special{pa 2880 360}%
\special{pa 2814 340}%
\special{pa 2828 360}%
\special{pa 2814 380}%
\special{pa 2880 360}%
\special{fp}%
% LINE 2 0 3 0
% 2 2600 840 2920 840
% 
\special{pn 8}%
\special{pa 2600 840}%
\special{pa 2920 840}%
\special{fp}%
% VECTOR 2 0 3 0
% 2 2920 840 2920 400
% 
\special{pn 8}%
\special{pa 2920 840}%
\special{pa 2920 400}%
\special{fp}%
\special{sh 1}%
\special{pa 2920 400}%
\special{pa 2900 468}%
\special{pa 2920 454}%
\special{pa 2940 468}%
\special{pa 2920 400}%
\special{fp}%
% CIRCLE 2 0 3 0
% 4 2920 360 2920 400 2920 400 2920 400
% 
\special{pn 8}%
\special{ar 2920 360 40 40  0.0000000 6.2831853}%
% VECTOR 2 0 3 0
% 2 2960 360 3280 360
% 
\special{pn 8}%
\special{pa 2960 360}%
\special{pa 3280 360}%
\special{fp}%
\special{sh 1}%
\special{pa 3280 360}%
\special{pa 3214 340}%
\special{pa 3228 360}%
\special{pa 3214 380}%
\special{pa 3280 360}%
\special{fp}%
% STR 2 0 3 0
% 3 2880 320 2880 400 4 0
% $+$
\put(28.8000,-4.0000){\makebox(0,0)[rt]{$+$}}%
% STR 2 0 3 0
% 3 1320 280 1320 360 5 0
% $\ds{3}$
\put(13.2000,-3.6000){\makebox(0,0){$\ds{3}$}}%
% STR 2 0 3 0
% 3 1800 280 1800 360 5 0
% $\us{3}$
\put(18.0000,-3.6000){\makebox(0,0){$\us{3}$}}%
% STR 2 0 3 0
% 3 2360 280 2360 360 5 0
% $\Phi_1(z)$
\put(23.6000,-3.6000){\makebox(0,0){$\Phi_1(z)$}}%
% STR 2 0 3 0
% 3 2360 760 2360 840 5 0
% $\Phi_2(z)$
\put(23.6000,-8.4000){\makebox(0,0){$\Phi_2(z)$}}%
% STR 2 0 3 0
% 3 1800 760 1800 840 5 0
% $\us{3}$
\put(18.0000,-8.4000){\makebox(0,0){$\us{3}$}}%
% STR 2 0 3 0
% 3 1320 760 1320 840 5 0
% $\ds{3}$
\put(13.2000,-8.4000){\makebox(0,0){$\ds{3}$}}%
% STR 2 0 3 0
% 3 840 760 840 840 5 0
% $z$
\put(8.4000,-8.4000){\makebox(0,0){$z$}}%
\end{picture}%

\end{center}
\caption{Nonuniform filterbank ($\M=[1,1,0]$).}
\label{fig:nonuniform-filterbank3}
\end{figure}

There is however another simpler way to implement the interpolation system,
by using a multirate filterbank, see Fig.\ \ref{fig:nonuniform-filterbank}.
In this filterbank, $\Phi_{i_1}(z)$, $\Phi_{i_2}(z)$, $\ldots$, $\Phi_{i_N}(z)$ 
are obtained by the following equation:
\[
  \left[\Phi_{i_1}(z),~\Phi_{i_2}(z),\ldots,\Phi_{i_N}(z)\right]
  = \left[1,~z^{-1},\ldots,z^{-M+1}\right]\tK(z^M).
\]
%The uniform decimator $\ds{M}$ and expander $\us{M}$ where $M$ is a positive integer is 
%defined by
%\[
%\begin{split}
%y &= (\ds{M})x,\quad 
%y_n = x_{Mn},\quad n=0,1,2,\ldots\\
%w &= (\us{M})v,\quad 
%w_n = \begin{cases} 
%		x_{n/M},\quad \text{if $n=0,M,2M,\ldots$,}\\
%		0,\quad \text{otherwise.}
%	\end{cases}
%\end{split}
%\]
%By using our definition of nonuniform decimator and expander,
%the uniform ones are given by
%\[
%\ds{M} = \ds{\M},\quad \us{M} = \us{\M}, \quad \M = [1, \underbrace{0,0,\ldots,0}_{M-1}].
%\]
Figure \ref{fig:nonuniform-filterbank3} shows an example of a nonuniform filterbank
when $\M = [1,1,0]$.
\section{Optimal Decimation Patterns}
\label{sec:optimal-pattern}
As shown in the previous section, 
if the decimation pattern $\M$ is given,
we can numerically find the $H^\infty$ optimal interpolation 
for non-band-limited signals in $\FL$
via the fast sampling method.
In this section, we consider designing the decimation pattern $\M$.

We observe that there exist several decimation patterns 
with the same decimation ratio $M/N$.
Consider $M=3$ and $N=2$. 
Then there are three patterns of decimation:
$\M_1 = [1,1,0]$, $\M_2 = [1,0,1]$, and $\M_3 = [0,1,1]$.
These are essentially the same except for one- or two-step delays,
that is,
\[
\begin{split}
z^{-1}(\us{\M_2})(\ds{\M_2}) &= (\us{\M_1})(\ds{\M_1})z^{-1},\\
(\us{\M_3})(\ds{\M_3})z^{-1} &= z^{-1}(\us{\M_1})(\ds{\M_1}).
\end{split}
\]
On the other hand, when $M=4$ and $N=2$, there can be a difference.
In this case, the essential patterns are
$\M=[1,1,0,0]$ and $\M=[1,0,1,0]$.
What is the difference between these two?

To see the difference, consider the following problem:
{\em find the optimal
decimation pattern(s) with the same ratio,
in view of the ability of signal reconstruction
for non-band-limited signals in $\FL$}.
More precisely, we formulate the problem as follows.
\begin{prob}
Given the decimation factors $M>0$ and $N>0$, find the optimal decimation pattern $\M$
which minimizes
\begin{equation}
\begin{split}
J(\M)
 &:=\min_{\tK}\|T_{ew}(\tK,\M)\|_\infty\\
 &= \min_{\tK}\sup_{\begin{subarray}{c}w\in L^2\\w \neq 0\end{subarray}}
 \frac{\|T_{ew}(\tK,\M)w\|_2}{\|w\|_2}.
\end{split}
\label{eq:optimal-pattern}
\end{equation}
\end{prob}

Since $\M$ is finite (that is, $M$ and $N$ are finite), this problem 
can be solved by optimizing (\ref{eq:hinf}) 
\[
 \binom{M}{N}=\frac{M!}{N!(M-N)!}
\]
times. 
%%% R2-9
Note that this is an upper bound of the number of optimization.
Counting the exact number is known as a necklace enumeration problem,
which is solved by so-called P\'{o}lya enumeration theorem
\cite{ZimFel06}.
%%%

\section{Design Examples}
\label{sec:examples}
In this section, we show design examples.
One can examine the simulation below by the MATLAB codes
provided in the web-page \cite{MAT}. 
\subsection{Optimal Filter Design}
We design the optimal filter $\tK$ (or $\Phi_{i_1}(z)$,\ldots $\Phi_{i_N}(z)$ in 
Fig.\ \ref{fig:nonuniform-filterbank}).
The design parameters are as follows:
the decimation pattern is $\M=[1,1,0]$, the sampling period $h=1$, 
the reconstruction delay $L=6$ (see Fig.~\ref{fig:errorsys}).
The transfer function of the original signal model $F(s)$ is set to be
\begin{equation}
F(s)=\frac{1}{10s+1}\label{eq:Fs}.
\end{equation}
%%% R2-1-2, R2-1-3
The fast-discretization ratio is empirically chosen as $n=4$,
which is sufficient for a good performance.

For comparison, we adopt the method of the Hilbert transformer \cite{VaiLiu90,Vai}
as a conventional one.
Note that this method is based on the sampling theorem, assuming that
the original analog signal is fully band-limited up to the frequency $\omega=2\pi/3$
($2/3$ of the Nyquist frequency $\pi$).
Note also that the conventional filter requires very large delay ($L=61.5$).

\begin{figure}[tb]
\begin{center}
%%% R2-12
\includegraphics[width=\linewidth]{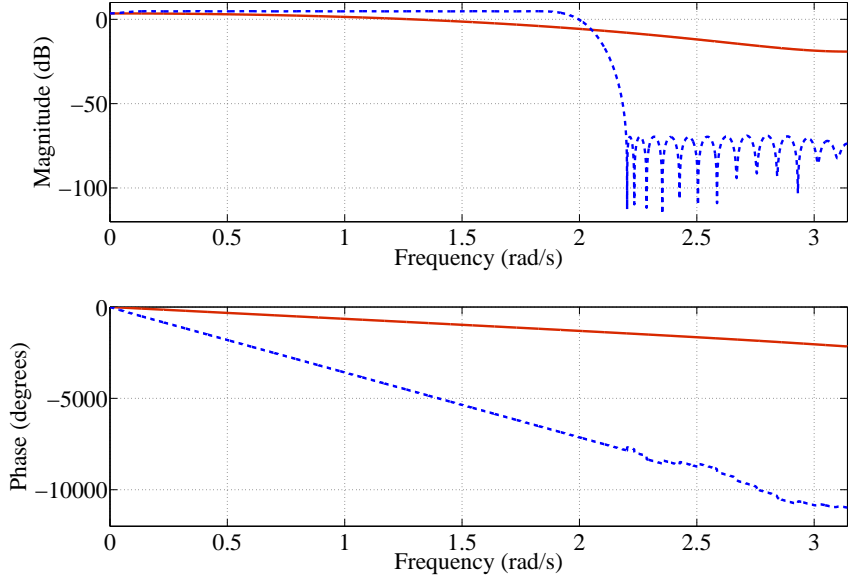}
\end{center}
\caption{Bode magnitude plot for the filter $\Phi_1(z)$ by $H^\infty$ optimal design (solid)
and a conventional design (dots).}
\label{fig:filters}
\end{figure}

\begin{figure}[tb]
\begin{center}
\includegraphics[width=\linewidth]{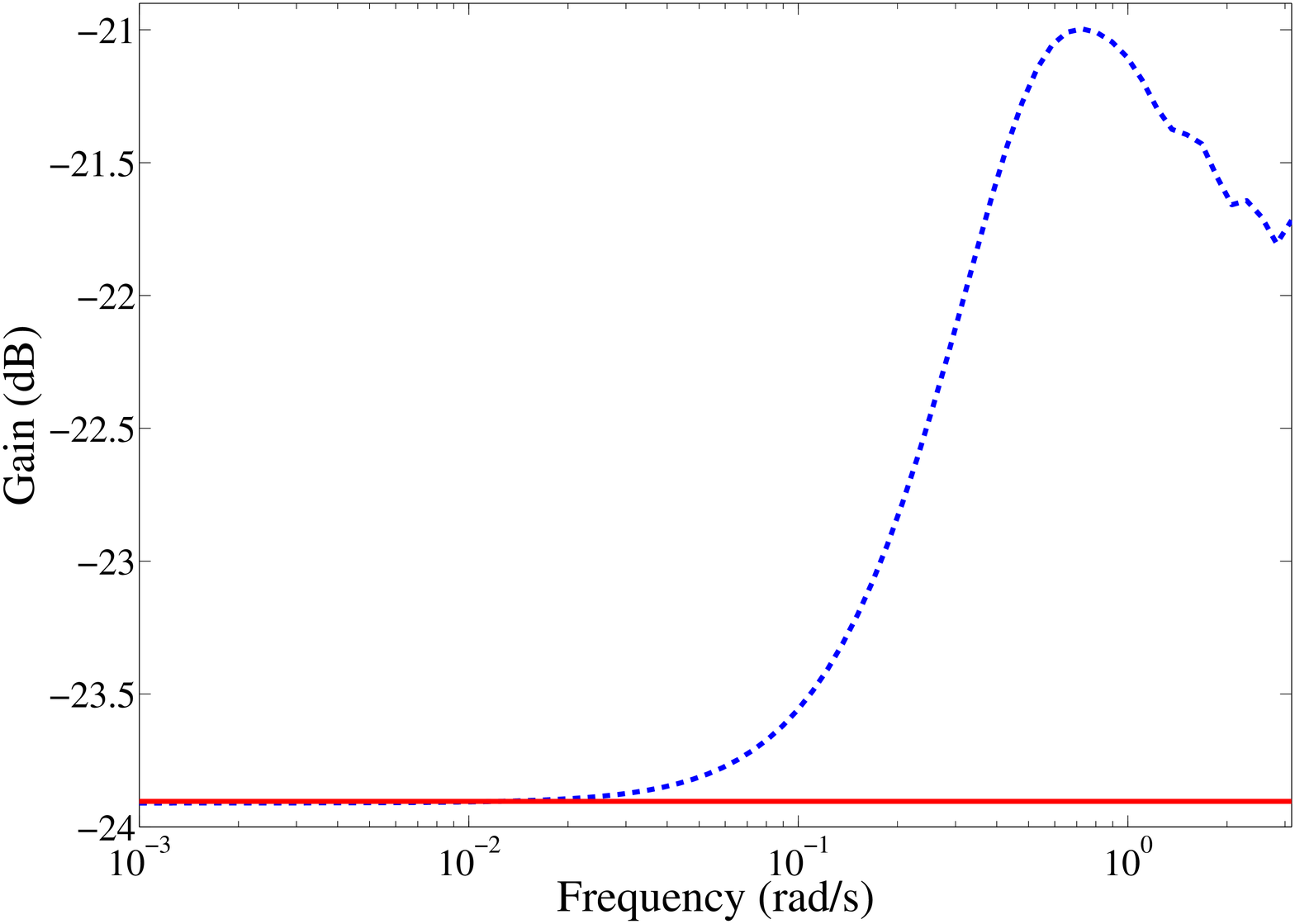}
\end{center}
\caption{Frequency response: proposed (solid) and conventional (dots).}
\label{fig:freqresp}
\end{figure}

Figure \ref{fig:filters} shows the Bode plots of the designed filter $\Phi_1(z)$
in the multirate filterbank implementation in Fig.~\ref{fig:FB}.
Since the conventional theory requires to perfectly cut off the frequency response
beyond the frequency $2/3\pi$ (rad/s),
the resulting filter shows a very sharp decay beyond this frequency.
On the other hand, our filter shows much slower decay.
To see this difference, we show in Fig.~\ref{fig:freqresp}
the frequency responses of the error system $T_{ew}(\tK,\M)$
in Fig.~\ref{fig:errorsys}.
The conventional interpolation shows a large error in high frequency,
while the sampled-data $H^\infty$ optimal interpolation shows a flat response.

\begin{figure}[tb]
\begin{center}
\includegraphics[width=\linewidth]{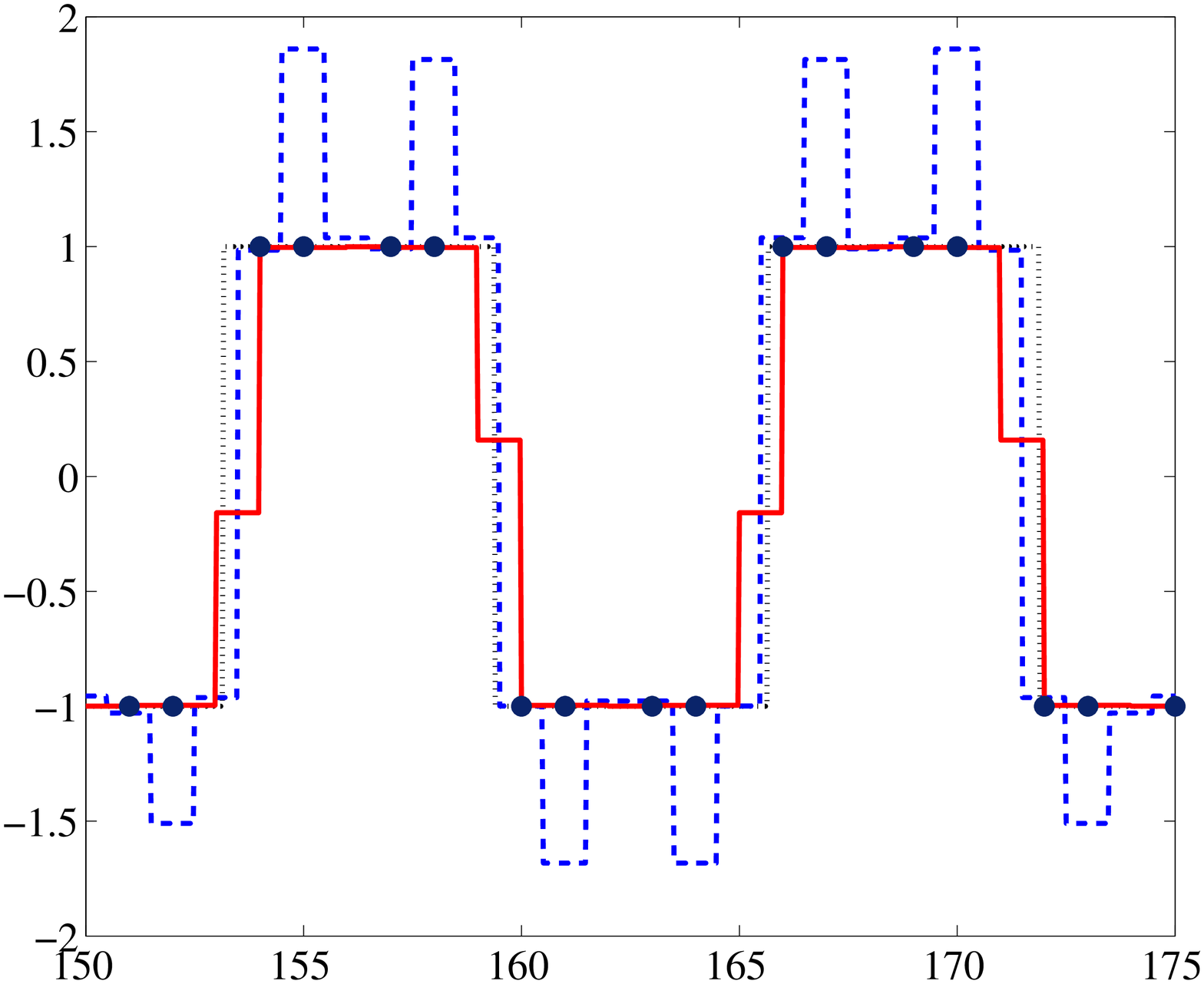}
\end{center}
\caption{Time response: proposed (solid), conventional (dash), and input signal(dots).}
\label{fig:timeresp}
\end{figure}

\begin{figure}[tb]
\begin{center}
\includegraphics[width=\linewidth]{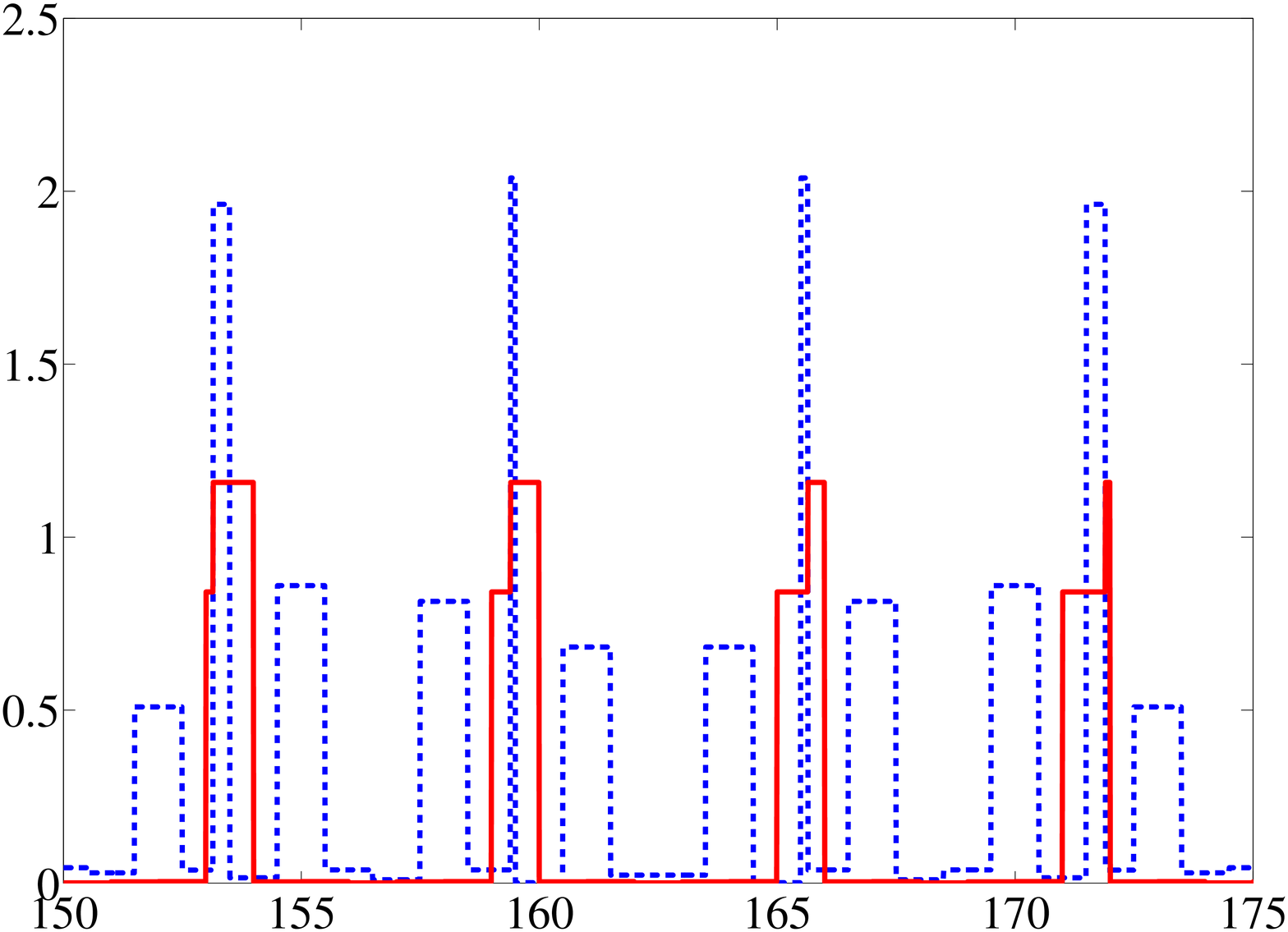}
\end{center}
\caption{Absolute Error in the time response in Fig.~\ref{fig:timeresp}: proposed (solid) and conventional (dash).}
\label{fig:abserrorresp}
\end{figure}

To illustrate the difference between these frequency responses,
we simulate interpolation of a rectangular wave.
Figure \ref{fig:timeresp} shows the time response.
The conventional interpolation causes large ripples,
while our interpolation shows a better response.
This is because the rectangular wave has high frequency components
around the edges, and our interpolation takes account of such 
frequency components.
Figure \ref{fig:abserrorresp} shows the absolute errors.
We can see that our response shows smaller errors than the conventional design.
\subsection{Decimation Pattern Analysis}
\label{subsec:ex-pattern}
Here we find the optimal decimation patterns
%for given decimation patterns $\M$.
for given decimation ratio $M/N$.
The sampling period is assumed to be $h=1$.
The transfer function $F(s)$ is given by (\ref{eq:Fs}).
The reconstruction delay is set $L=M$ (the length of $\M$).

First, Let $M=4$ and $N=2$.
In this case, the essential patterns are
$\M=[1,1,0,0]$ and $\M=[1,0,1,0]$.
Table\ \ref{tbl:M4N2} shows the optimal value $J(\M)$ 
defined in (\ref{eq:optimal-pattern}).
\begin{table}[tb]
\caption{Optimal value $J(\M)$ for $M=4$ and $N=2$}
\label{tbl:M4N2}
\begin{center}
\begin{tabular}{|c|c|}\hline
Decimation Pattern   &  $J(\M)$ \\\hline
					&  \\
%%%%\input{1100s.tex}
%WinTpicVersion3.08
\unitlength 0.1in
\begin{picture}(  4.0000,  2.5000)(  3.5000, -6.0000)
% LINE 2 0 3 0
% 2 375 575 375 400
% 
\special{pn 8}%
\special{pa 376 576}%
\special{pa 376 400}%
\special{fp}%
% CIRCLE 2 0 0 0
% 4 375 375 375 400 375 400 375 400
% 
\special{pn 8}%
\special{sh 0.600}%
\special{ar 376 376 26 26  0.0000000 6.2831853}%
% LINE 2 0 3 0
% 2 375 575 550 575
% 
\special{pn 8}%
\special{pa 376 576}%
\special{pa 550 576}%
\special{fp}%
% LINE 2 0 3 0
% 2 475 575 475 400
% 
\special{pn 8}%
\special{pa 476 576}%
\special{pa 476 400}%
\special{fp}%
% CIRCLE 2 0 0 0
% 4 475 375 475 400 475 400 475 400
% 
\special{pn 8}%
\special{sh 0.600}%
\special{ar 476 376 26 26  0.0000000 6.2831853}%
% CIRCLE 2 0 3 0
% 4 575 575 575 600 575 600 575 600
% 
\special{pn 8}%
\special{ar 576 576 26 26  0.0000000 6.2831853}%
% LINE 2 0 3 0
% 4 600 575 650 575 650 575 650 575
% 
\special{pn 8}%
\special{pa 600 576}%
\special{pa 650 576}%
\special{fp}%
\special{pa 650 576}%
\special{pa 650 576}%
\special{fp}%
% CIRCLE 2 0 3 0
% 4 675 575 675 600 675 600 675 600
% 
\special{pn 8}%
\special{ar 676 576 26 26  0.0000000 6.2831853}%
% LINE 2 0 3 0
% 2 700 575 750 575
% 
\special{pn 8}%
\special{pa 700 576}%
\special{pa 750 576}%
\special{fp}%
\end{picture}%

 & 0.2293\\
$\M = [1 1 0 0]$, $[1 0 0 1]$, $[0 1 1 0]$, $[0 0 1 1]$ & \\\hline
				&  \\
%%%%\input{1010s.tex}
%WinTpicVersion3.08
\unitlength 0.1in
\begin{picture}(  4.0000,  2.5000)(  3.5000, -6.0000)
% LINE 2 0 3 0
% 2 375 575 375 400
% 
\special{pn 8}%
\special{pa 376 576}%
\special{pa 376 400}%
\special{fp}%
% CIRCLE 2 0 0 0
% 4 375 375 375 400 375 400 375 400
% 
\special{pn 8}%
\special{sh 0.600}%
\special{ar 376 376 26 26  0.0000000 6.2831853}%
% LINE 2 0 3 0
% 2 575 575 575 400
% 
\special{pn 8}%
\special{pa 576 576}%
\special{pa 576 400}%
\special{fp}%
% CIRCLE 2 0 0 0
% 4 575 375 575 400 575 400 575 400
% 
\special{pn 8}%
\special{sh 0.600}%
\special{ar 576 376 26 26  0.0000000 6.2831853}%
% CIRCLE 2 0 3 0
% 4 475 575 475 600 475 600 475 600
% 
\special{pn 8}%
\special{ar 476 576 26 26  0.0000000 6.2831853}%
% LINE 2 0 3 0
% 4 600 575 650 575 650 575 650 575
% 
\special{pn 8}%
\special{pa 600 576}%
\special{pa 650 576}%
\special{fp}%
\special{pa 650 576}%
\special{pa 650 576}%
\special{fp}%
% CIRCLE 2 0 3 0
% 4 675 575 675 600 675 600 675 600
% 
\special{pn 8}%
\special{ar 676 576 26 26  0.0000000 6.2831853}%
% LINE 2 0 3 0
% 2 700 575 750 575
% 
\special{pn 8}%
\special{pa 700 576}%
\special{pa 750 576}%
\special{fp}%
% LINE 2 0 3 0
% 2 375 575 450 575
% 
\special{pn 8}%
\special{pa 376 576}%
\special{pa 450 576}%
\special{fp}%
% LINE 2 0 3 0
% 2 500 575 600 575
% 
\special{pn 8}%
\special{pa 500 576}%
\special{pa 600 576}%
\special{fp}%
\end{picture}%

 & 0.1529\\
$\M = [1 0 1 0]$,  $[0 1 0 1]$ & \\\hline
\end{tabular}
\end{center}
\end{table}
We can see the difference between the two patterns
with the same decimation ratio.
This result shows that 
the pattern $\M=[1,0,1,0]$ (or $\M=[0,1,0,1]$) is
the better,
which is equal to the uniform decimation $\ds{2}$.

We then consider when the segment length $M=5$.
Table\ \ref{tbl:M5} shows the result.
By this, the optimal value $J(\M)$ depends on
the position of the zeros in $\M$,
not depends on the number of the ones in $\M$.
For example, although the pattern (F) retains more samples
than the pattern (E), the optimal values are the same.
This fact shows that a lower ratio of decimation (or compression)
does not necessarily lead to a better performance.
In other words, not only decimation ratio but also decimation pattern plays an important role in signal compression.
\begin{table}[tb]
\caption{Optimal value $J(\M)$ for $M=5$ and $N=1,2,3,4$}
\label{tbl:M5}
\begin{center}
\begin{tabular}{|c|c||c|c|}\hline
Pattern   &  $J(\M)$ &  Pattern   &  $J(\M)$ \\\hline
%$N=1$					&  & $N=3$					&  \\
 & & & \\
%%%%\input{10000s.tex} 
%WinTpicVersion3.08
\unitlength 0.1in
\begin{picture}(  5.0000,  2.5000)(  3.5000, -6.0000)
% LINE 2 0 3 0
% 2 375 575 375 400
% 
\special{pn 8}%
\special{pa 376 576}%
\special{pa 376 400}%
\special{fp}%
% CIRCLE 2 0 0 0
% 4 375 375 375 400 375 400 375 400
% 
\special{pn 8}%
\special{sh 0.600}%
\special{ar 376 376 26 26  0.0000000 6.2831853}%
% CIRCLE 2 0 3 0
% 4 575 575 575 600 575 600 575 600
% 
\special{pn 8}%
\special{ar 576 576 26 26  0.0000000 6.2831853}%
% CIRCLE 2 0 3 0
% 4 775 575 775 600 775 600 775 600
% 
\special{pn 8}%
\special{ar 776 576 26 26  0.0000000 6.2831853}%
% LINE 2 0 3 0
% 2 800 575 850 575
% 
\special{pn 8}%
\special{pa 800 576}%
\special{pa 850 576}%
\special{fp}%
% LINE 2 0 3 0
% 2 700 575 750 575
% 
\special{pn 8}%
\special{pa 700 576}%
\special{pa 750 576}%
\special{fp}%
% CIRCLE 2 0 3 0
% 4 675 575 675 600 675 600 675 600
% 
\special{pn 8}%
\special{ar 676 576 26 26  0.0000000 6.2831853}%
% LINE 2 0 3 0
% 2 650 575 600 575
% 
\special{pn 8}%
\special{pa 650 576}%
\special{pa 600 576}%
\special{fp}%
% LINE 2 0 3 0
% 2 375 575 450 575
% 
\special{pn 8}%
\special{pa 376 576}%
\special{pa 450 576}%
\special{fp}%
% CIRCLE 2 0 3 0
% 4 475 575 475 600 475 600 475 600
% 
\special{pn 8}%
\special{ar 476 576 26 26  0.0000000 6.2831853}%
% LINE 2 0 3 0
% 2 500 575 550 575
% 
\special{pn 8}%
\special{pa 500 576}%
\special{pa 550 576}%
\special{fp}%
\end{picture}%

& 0.3813 &
%%%%\input{11100s.tex}
%WinTpicVersion3.08
\unitlength 0.1in
\begin{picture}(  5.0000,  2.5000)(  3.5000, -6.0000)
% LINE 2 0 3 0
% 2 375 575 375 400
% 
\special{pn 8}%
\special{pa 376 576}%
\special{pa 376 400}%
\special{fp}%
% CIRCLE 2 0 0 0
% 4 375 375 375 400 375 400 375 400
% 
\special{pn 8}%
\special{sh 0.600}%
\special{ar 376 376 26 26  0.0000000 6.2831853}%
% LINE 2 0 3 0
% 2 375 575 550 575
% 
\special{pn 8}%
\special{pa 376 576}%
\special{pa 550 576}%
\special{fp}%
% LINE 2 0 3 0
% 2 475 575 475 400
% 
\special{pn 8}%
\special{pa 476 576}%
\special{pa 476 400}%
\special{fp}%
% CIRCLE 2 0 0 0
% 4 475 375 475 400 475 400 475 400
% 
\special{pn 8}%
\special{sh 0.600}%
\special{ar 476 376 26 26  0.0000000 6.2831853}%
% CIRCLE 2 0 3 0
% 4 675 575 675 600 675 600 675 600
% 
\special{pn 8}%
\special{ar 676 576 26 26  0.0000000 6.2831853}%
% LINE 2 0 3 0
% 4 700 575 750 575 750 575 750 575
% 
\special{pn 8}%
\special{pa 700 576}%
\special{pa 750 576}%
\special{fp}%
\special{pa 750 576}%
\special{pa 750 576}%
\special{fp}%
% CIRCLE 2 0 3 0
% 4 775 575 775 600 775 600 775 600
% 
\special{pn 8}%
\special{ar 776 576 26 26  0.0000000 6.2831853}%
% LINE 2 0 3 0
% 2 800 575 850 575
% 
\special{pn 8}%
\special{pa 800 576}%
\special{pa 850 576}%
\special{fp}%
% LINE 2 0 3 0
% 2 550 575 650 575
% 
\special{pn 8}%
\special{pa 550 576}%
\special{pa 650 576}%
\special{fp}%
% LINE 2 0 3 0
% 2 575 575 575 400
% 
\special{pn 8}%
\special{pa 576 576}%
\special{pa 576 400}%
\special{fp}%
% CIRCLE 2 0 0 0
% 4 575 375 575 400 575 400 575 400
% 
\special{pn 8}%
\special{sh 0.600}%
\special{ar 576 376 26 26  0.0000000 6.2831853}%
\end{picture}%

 & 0.2303\\
(A) & &(D) & \\\cline{1-2}
% & \\\hline
%$N=2$					&  &\input{11010s.tex} & 0.1536\\
 & & & \\
%%%%\input{11000s.tex}
%WinTpicVersion3.08
\unitlength 0.1in
\begin{picture}(  5.0000,  2.5000)(  3.5000, -6.0000)
% LINE 2 0 3 0
% 2 375 575 375 400
% 
\special{pn 8}%
\special{pa 376 576}%
\special{pa 376 400}%
\special{fp}%
% CIRCLE 2 0 0 0
% 4 375 375 375 400 375 400 375 400
% 
\special{pn 8}%
\special{sh 0.600}%
\special{ar 376 376 26 26  0.0000000 6.2831853}%
% LINE 2 0 3 0
% 2 375 575 550 575
% 
\special{pn 8}%
\special{pa 376 576}%
\special{pa 550 576}%
\special{fp}%
% LINE 2 0 3 0
% 2 475 575 475 400
% 
\special{pn 8}%
\special{pa 476 576}%
\special{pa 476 400}%
\special{fp}%
% CIRCLE 2 0 0 0
% 4 475 375 475 400 475 400 475 400
% 
\special{pn 8}%
\special{sh 0.600}%
\special{ar 476 376 26 26  0.0000000 6.2831853}%
% CIRCLE 2 0 3 0
% 4 575 575 575 600 575 600 575 600
% 
\special{pn 8}%
\special{ar 576 576 26 26  0.0000000 6.2831853}%
% CIRCLE 2 0 3 0
% 4 775 575 775 600 775 600 775 600
% 
\special{pn 8}%
\special{ar 776 576 26 26  0.0000000 6.2831853}%
% LINE 2 0 3 0
% 2 800 575 850 575
% 
\special{pn 8}%
\special{pa 800 576}%
\special{pa 850 576}%
\special{fp}%
% LINE 2 0 3 0
% 2 700 575 750 575
% 
\special{pn 8}%
\special{pa 700 576}%
\special{pa 750 576}%
\special{fp}%
% CIRCLE 2 0 3 0
% 4 675 575 675 600 675 600 675 600
% 
\special{pn 8}%
\special{ar 676 576 26 26  0.0000000 6.2831853}%
% LINE 2 0 3 0
% 2 650 575 600 575
% 
\special{pn 8}%
\special{pa 650 576}%
\special{pa 600 576}%
\special{fp}%
\end{picture}%

 & 0.3062 &
%%%%\input{11010s.tex}
%WinTpicVersion3.08
\unitlength 0.1in
\begin{picture}(  5.0000,  2.5000)(  3.5000, -6.0000)
% LINE 2 0 3 0
% 2 375 575 375 400
% 
\special{pn 8}%
\special{pa 376 576}%
\special{pa 376 400}%
\special{fp}%
% CIRCLE 2 0 0 0
% 4 375 375 375 400 375 400 375 400
% 
\special{pn 8}%
\special{sh 0.600}%
\special{ar 376 376 26 26  0.0000000 6.2831853}%
% LINE 2 0 3 0
% 2 375 575 550 575
% 
\special{pn 8}%
\special{pa 376 576}%
\special{pa 550 576}%
\special{fp}%
% LINE 2 0 3 0
% 2 475 575 475 400
% 
\special{pn 8}%
\special{pa 476 576}%
\special{pa 476 400}%
\special{fp}%
% CIRCLE 2 0 0 0
% 4 475 375 475 400 475 400 475 400
% 
\special{pn 8}%
\special{sh 0.600}%
\special{ar 476 376 26 26  0.0000000 6.2831853}%
% CIRCLE 2 0 3 0
% 4 575 575 575 600 575 600 575 600
% 
\special{pn 8}%
\special{ar 576 576 26 26  0.0000000 6.2831853}%
% LINE 2 0 3 0
% 4 700 575 750 575 750 575 750 575
% 
\special{pn 8}%
\special{pa 700 576}%
\special{pa 750 576}%
\special{fp}%
\special{pa 750 576}%
\special{pa 750 576}%
\special{fp}%
% CIRCLE 2 0 3 0
% 4 775 575 775 600 775 600 775 600
% 
\special{pn 8}%
\special{ar 776 576 26 26  0.0000000 6.2831853}%
% LINE 2 0 3 0
% 2 800 575 850 575
% 
\special{pn 8}%
\special{pa 800 576}%
\special{pa 850 576}%
\special{fp}%
% LINE 2 0 3 0
% 2 675 575 675 400
% 
\special{pn 8}%
\special{pa 676 576}%
\special{pa 676 400}%
\special{fp}%
% CIRCLE 2 0 0 0
% 4 675 375 675 400 675 400 675 400
% 
\special{pn 8}%
\special{sh 0.600}%
\special{ar 676 376 26 26  0.0000000 6.2831853}%
% LINE 2 0 3 0
% 2 600 575 700 575
% 
\special{pn 8}%
\special{pa 600 576}%
\special{pa 700 576}%
\special{fp}%
\end{picture}%

 & 0.1536\\
(B) & &(E) & \\\cline{3-4}
%(B) & &$N=4$					&  \\
 & & & \\
%					&  \\
%%%%\input{10100s.tex}
%WinTpicVersion3.08
\unitlength 0.1in
\begin{picture}(  5.0000,  2.5000)(  3.5000, -6.0000)
% LINE 2 0 3 0
% 2 375 575 375 400
% 
\special{pn 8}%
\special{pa 376 576}%
\special{pa 376 400}%
\special{fp}%
% CIRCLE 2 0 0 0
% 4 375 375 375 400 375 400 375 400
% 
\special{pn 8}%
\special{sh 0.600}%
\special{ar 376 376 26 26  0.0000000 6.2831853}%
% LINE 2 0 3 0
% 2 575 575 575 400
% 
\special{pn 8}%
\special{pa 576 576}%
\special{pa 576 400}%
\special{fp}%
% CIRCLE 2 0 0 0
% 4 575 375 575 400 575 400 575 400
% 
\special{pn 8}%
\special{sh 0.600}%
\special{ar 576 376 26 26  0.0000000 6.2831853}%
% CIRCLE 2 0 3 0
% 4 475 575 475 600 475 600 475 600
% 
\special{pn 8}%
\special{ar 476 576 26 26  0.0000000 6.2831853}%
% CIRCLE 2 0 3 0
% 4 775 575 775 600 775 600 775 600
% 
\special{pn 8}%
\special{ar 776 576 26 26  0.0000000 6.2831853}%
% LINE 2 0 3 0
% 2 800 575 850 575
% 
\special{pn 8}%
\special{pa 800 576}%
\special{pa 850 576}%
\special{fp}%
% LINE 2 0 3 0
% 2 700 575 750 575
% 
\special{pn 8}%
\special{pa 700 576}%
\special{pa 750 576}%
\special{fp}%
% CIRCLE 2 0 3 0
% 4 675 575 675 600 675 600 675 600
% 
\special{pn 8}%
\special{ar 676 576 26 26  0.0000000 6.2831853}%
% LINE 2 0 3 0
% 2 650 575 600 575
% 
\special{pn 8}%
\special{pa 650 576}%
\special{pa 600 576}%
\special{fp}%
% LINE 2 0 3 0
% 2 375 575 450 575
% 
\special{pn 8}%
\special{pa 376 576}%
\special{pa 450 576}%
\special{fp}%
% LINE 2 0 3 0
% 2 500 575 600 575
% 
\special{pn 8}%
\special{pa 500 576}%
\special{pa 600 576}%
\special{fp}%
\end{picture}%

 & 0.2303 &
%%%%\input{11110s.tex}
%WinTpicVersion3.08
\unitlength 0.1in
\begin{picture}(  5.0000,  2.5000)(  3.5000, -6.0000)
% LINE 2 0 3 0
% 2 375 575 375 400
% 
\special{pn 8}%
\special{pa 376 576}%
\special{pa 376 400}%
\special{fp}%
% CIRCLE 2 0 0 0
% 4 375 375 375 400 375 400 375 400
% 
\special{pn 8}%
\special{sh 0.600}%
\special{ar 376 376 26 26  0.0000000 6.2831853}%
% LINE 2 0 3 0
% 2 375 575 550 575
% 
\special{pn 8}%
\special{pa 376 576}%
\special{pa 550 576}%
\special{fp}%
% LINE 2 0 3 0
% 2 475 575 475 400
% 
\special{pn 8}%
\special{pa 476 576}%
\special{pa 476 400}%
\special{fp}%
% CIRCLE 2 0 0 0
% 4 475 375 475 400 475 400 475 400
% 
\special{pn 8}%
\special{sh 0.600}%
\special{ar 476 376 26 26  0.0000000 6.2831853}%
% LINE 2 0 3 0
% 4 700 575 750 575 750 575 750 575
% 
\special{pn 8}%
\special{pa 700 576}%
\special{pa 750 576}%
\special{fp}%
\special{pa 750 576}%
\special{pa 750 576}%
\special{fp}%
% CIRCLE 2 0 3 0
% 4 775 575 775 600 775 600 775 600
% 
\special{pn 8}%
\special{ar 776 576 26 26  0.0000000 6.2831853}%
% LINE 2 0 3 0
% 2 800 575 850 575
% 
\special{pn 8}%
\special{pa 800 576}%
\special{pa 850 576}%
\special{fp}%
% LINE 2 0 3 0
% 2 550 575 650 575
% 
\special{pn 8}%
\special{pa 550 576}%
\special{pa 650 576}%
\special{fp}%
% LINE 2 0 3 0
% 2 575 575 575 400
% 
\special{pn 8}%
\special{pa 576 576}%
\special{pa 576 400}%
\special{fp}%
% CIRCLE 2 0 0 0
% 4 575 375 575 400 575 400 575 400
% 
\special{pn 8}%
\special{sh 0.600}%
\special{ar 576 376 26 26  0.0000000 6.2831853}%
% LINE 2 0 3 0
% 2 675 575 675 400
% 
\special{pn 8}%
\special{pa 676 576}%
\special{pa 676 400}%
\special{fp}%
% CIRCLE 2 0 0 0
% 4 675 375 675 400 675 400 675 400
% 
\special{pn 8}%
\special{sh 0.600}%
\special{ar 676 376 26 26  0.0000000 6.2831853}%
% LINE 2 0 3 0
% 2 650 575 700 575
% 
\special{pn 8}%
\special{pa 650 576}%
\special{pa 700 576}%
\special{fp}%
\end{picture}%

 & 0.1536\\
(C) & &(F) & \\\hline
\end{tabular}
\end{center}
\end{table}

Table\ \ref{tbl:M7N4} shows the optimal $J(\M)$
when $M=7$ and $N=4$, that is, the decimation ratio is $7/4$.
\begin{table}[tb]
\caption{Optimal value $J(\M)$ for $M=7$, $N=4$.}
\label{tbl:M7N4}
\begin{center}
\begin{tabular}{|c|c|}\hline
Decimation Pattern   &  $J(\M)$ \\\hline
					&  \\
%%%%\input{1111000s.tex}
%WinTpicVersion3.08
\unitlength 0.1in
\begin{picture}(  7.0000,  2.5000)(  3.5000, -6.0000)
% LINE 2 0 3 0
% 2 375 575 375 400
% 
\special{pn 8}%
\special{pa 376 576}%
\special{pa 376 400}%
\special{fp}%
% CIRCLE 2 0 0 0
% 4 375 375 375 400 375 400 375 400
% 
\special{pn 8}%
\special{sh 0.600}%
\special{ar 376 376 26 26  0.0000000 6.2831853}%
% LINE 2 0 3 0
% 2 375 575 550 575
% 
\special{pn 8}%
\special{pa 376 576}%
\special{pa 550 576}%
\special{fp}%
% LINE 2 0 3 0
% 2 475 575 475 400
% 
\special{pn 8}%
\special{pa 476 576}%
\special{pa 476 400}%
\special{fp}%
% CIRCLE 2 0 0 0
% 4 475 375 475 400 475 400 475 400
% 
\special{pn 8}%
\special{sh 0.600}%
\special{ar 476 376 26 26  0.0000000 6.2831853}%
% LINE 2 0 3 0
% 4 700 575 750 575 750 575 750 575
% 
\special{pn 8}%
\special{pa 700 576}%
\special{pa 750 576}%
\special{fp}%
\special{pa 750 576}%
\special{pa 750 576}%
\special{fp}%
% CIRCLE 2 0 3 0
% 4 775 575 775 600 775 600 775 600
% 
\special{pn 8}%
\special{ar 776 576 26 26  0.0000000 6.2831853}%
% LINE 2 0 3 0
% 2 800 575 850 575
% 
\special{pn 8}%
\special{pa 800 576}%
\special{pa 850 576}%
\special{fp}%
% LINE 2 0 3 0
% 2 550 575 650 575
% 
\special{pn 8}%
\special{pa 550 576}%
\special{pa 650 576}%
\special{fp}%
% LINE 2 0 3 0
% 2 575 575 575 400
% 
\special{pn 8}%
\special{pa 576 576}%
\special{pa 576 400}%
\special{fp}%
% CIRCLE 2 0 0 0
% 4 575 375 575 400 575 400 575 400
% 
\special{pn 8}%
\special{sh 0.600}%
\special{ar 576 376 26 26  0.0000000 6.2831853}%
% LINE 2 0 3 0
% 2 675 575 675 400
% 
\special{pn 8}%
\special{pa 676 576}%
\special{pa 676 400}%
\special{fp}%
% CIRCLE 2 0 0 0
% 4 675 375 675 400 675 400 675 400
% 
\special{pn 8}%
\special{sh 0.600}%
\special{ar 676 376 26 26  0.0000000 6.2831853}%
% LINE 2 0 3 0
% 2 650 575 700 575
% 
\special{pn 8}%
\special{pa 650 576}%
\special{pa 700 576}%
\special{fp}%
% CIRCLE 2 0 3 0
% 4 875 575 875 600 875 600 875 600
% 
\special{pn 8}%
\special{ar 876 576 26 26  0.0000000 6.2831853}%
% CIRCLE 2 0 3 0
% 4 975 575 975 600 975 600 975 600
% 
\special{pn 8}%
\special{ar 976 576 26 26  0.0000000 6.2831853}%
% LINE 2 0 3 0
% 2 900 575 950 575
% 
\special{pn 8}%
\special{pa 900 576}%
\special{pa 950 576}%
\special{fp}%
% LINE 2 0 3 0
% 2 1000 575 1050 575
% 
\special{pn 8}%
\special{pa 1000 576}%
\special{pa 1050 576}%
\special{fp}%
\end{picture}%

 & 0.3089\\
(A)					&  \\\hline
 & \\
%%%%\input{1110100s.tex}
%WinTpicVersion3.08
\unitlength 0.1in
\begin{picture}(  7.0000,  2.5000)(  3.5000, -6.0000)
% LINE 2 0 3 0
% 2 375 575 375 400
% 
\special{pn 8}%
\special{pa 376 576}%
\special{pa 376 400}%
\special{fp}%
% CIRCLE 2 0 0 0
% 4 375 375 375 400 375 400 375 400
% 
\special{pn 8}%
\special{sh 0.600}%
\special{ar 376 376 26 26  0.0000000 6.2831853}%
% LINE 2 0 3 0
% 2 375 575 550 575
% 
\special{pn 8}%
\special{pa 376 576}%
\special{pa 550 576}%
\special{fp}%
% LINE 2 0 3 0
% 2 475 575 475 400
% 
\special{pn 8}%
\special{pa 476 576}%
\special{pa 476 400}%
\special{fp}%
% CIRCLE 2 0 0 0
% 4 475 375 475 400 475 400 475 400
% 
\special{pn 8}%
\special{sh 0.600}%
\special{ar 476 376 26 26  0.0000000 6.2831853}%
% LINE 2 0 3 0
% 4 700 575 750 575 750 575 750 575
% 
\special{pn 8}%
\special{pa 700 576}%
\special{pa 750 576}%
\special{fp}%
\special{pa 750 576}%
\special{pa 750 576}%
\special{fp}%
% CIRCLE 2 0 3 0
% 4 675 575 675 600 675 600 675 600
% 
\special{pn 8}%
\special{ar 676 576 26 26  0.0000000 6.2831853}%
% LINE 2 0 3 0
% 2 800 575 850 575
% 
\special{pn 8}%
\special{pa 800 576}%
\special{pa 850 576}%
\special{fp}%
% LINE 2 0 3 0
% 2 550 575 650 575
% 
\special{pn 8}%
\special{pa 550 576}%
\special{pa 650 576}%
\special{fp}%
% LINE 2 0 3 0
% 2 575 575 575 400
% 
\special{pn 8}%
\special{pa 576 576}%
\special{pa 576 400}%
\special{fp}%
% CIRCLE 2 0 0 0
% 4 575 375 575 400 575 400 575 400
% 
\special{pn 8}%
\special{sh 0.600}%
\special{ar 576 376 26 26  0.0000000 6.2831853}%
% LINE 2 0 3 0
% 2 775 575 775 400
% 
\special{pn 8}%
\special{pa 776 576}%
\special{pa 776 400}%
\special{fp}%
% CIRCLE 2 0 0 0
% 4 775 375 775 400 775 400 775 400
% 
\special{pn 8}%
\special{sh 0.600}%
\special{ar 776 376 26 26  0.0000000 6.2831853}%
% CIRCLE 2 0 3 0
% 4 875 575 875 600 875 600 875 600
% 
\special{pn 8}%
\special{ar 876 576 26 26  0.0000000 6.2831853}%
% CIRCLE 2 0 3 0
% 4 975 575 975 600 975 600 975 600
% 
\special{pn 8}%
\special{ar 976 576 26 26  0.0000000 6.2831853}%
% LINE 2 0 3 0
% 2 900 575 950 575
% 
\special{pn 8}%
\special{pa 900 576}%
\special{pa 950 576}%
\special{fp}%
% LINE 2 0 3 0
% 2 1000 575 1050 575
% 
\special{pn 8}%
\special{pa 1000 576}%
\special{pa 1050 576}%
\special{fp}%
% LINE 2 0 3 0
% 2 750 575 800 575
% 
\special{pn 8}%
\special{pa 750 576}%
\special{pa 800 576}%
\special{fp}%
\end{picture}%

 & \\
(B)					&  \\
%%%%\input{1110010s.tex}
%WinTpicVersion3.08
\unitlength 0.1in
\begin{picture}(  7.0000,  2.5000)(  3.5000, -6.0000)
% LINE 2 0 3 0
% 2 375 575 375 400
% 
\special{pn 8}%
\special{pa 376 576}%
\special{pa 376 400}%
\special{fp}%
% CIRCLE 2 0 0 0
% 4 375 375 375 400 375 400 375 400
% 
\special{pn 8}%
\special{sh 0.600}%
\special{ar 376 376 26 26  0.0000000 6.2831853}%
% LINE 2 0 3 0
% 2 375 575 550 575
% 
\special{pn 8}%
\special{pa 376 576}%
\special{pa 550 576}%
\special{fp}%
% LINE 2 0 3 0
% 2 475 575 475 400
% 
\special{pn 8}%
\special{pa 476 576}%
\special{pa 476 400}%
\special{fp}%
% CIRCLE 2 0 0 0
% 4 475 375 475 400 475 400 475 400
% 
\special{pn 8}%
\special{sh 0.600}%
\special{ar 476 376 26 26  0.0000000 6.2831853}%
% LINE 2 0 3 0
% 4 700 575 750 575 750 575 750 575
% 
\special{pn 8}%
\special{pa 700 576}%
\special{pa 750 576}%
\special{fp}%
\special{pa 750 576}%
\special{pa 750 576}%
\special{fp}%
% CIRCLE 2 0 3 0
% 4 675 575 675 600 675 600 675 600
% 
\special{pn 8}%
\special{ar 676 576 26 26  0.0000000 6.2831853}%
% LINE 2 0 3 0
% 2 800 575 850 575
% 
\special{pn 8}%
\special{pa 800 576}%
\special{pa 850 576}%
\special{fp}%
% LINE 2 0 3 0
% 2 550 575 650 575
% 
\special{pn 8}%
\special{pa 550 576}%
\special{pa 650 576}%
\special{fp}%
% LINE 2 0 3 0
% 2 575 575 575 400
% 
\special{pn 8}%
\special{pa 576 576}%
\special{pa 576 400}%
\special{fp}%
% CIRCLE 2 0 0 0
% 4 575 375 575 400 575 400 575 400
% 
\special{pn 8}%
\special{sh 0.600}%
\special{ar 576 376 26 26  0.0000000 6.2831853}%
% LINE 2 0 3 0
% 2 875 575 875 400
% 
\special{pn 8}%
\special{pa 876 576}%
\special{pa 876 400}%
\special{fp}%
% CIRCLE 2 0 0 0
% 4 875 375 875 400 875 400 875 400
% 
\special{pn 8}%
\special{sh 0.600}%
\special{ar 876 376 26 26  0.0000000 6.2831853}%
% CIRCLE 2 0 3 0
% 4 775 575 775 600 775 600 775 600
% 
\special{pn 8}%
\special{ar 776 576 26 26  0.0000000 6.2831853}%
% CIRCLE 2 0 3 0
% 4 975 575 975 600 975 600 975 600
% 
\special{pn 8}%
\special{ar 976 576 26 26  0.0000000 6.2831853}%
% LINE 2 0 3 0
% 2 900 575 950 575
% 
\special{pn 8}%
\special{pa 900 576}%
\special{pa 950 576}%
\special{fp}%
% LINE 2 0 3 0
% 2 1000 575 1050 575
% 
\special{pn 8}%
\special{pa 1000 576}%
\special{pa 1050 576}%
\special{fp}%
% LINE 2 0 3 0
% 2 850 575 900 575
% 
\special{pn 8}%
\special{pa 850 576}%
\special{pa 900 576}%
\special{fp}%
\end{picture}%

 & 0.2320\\
(C)					&  \\
%%%%\input{1100110s.tex}
%WinTpicVersion3.08
\unitlength 0.1in
\begin{picture}(  7.0000,  2.5000)(  3.5000, -6.0000)
% LINE 2 0 3 0
% 2 375 575 375 400
% 
\special{pn 8}%
\special{pa 376 576}%
\special{pa 376 400}%
\special{fp}%
% CIRCLE 2 0 0 0
% 4 375 375 375 400 375 400 375 400
% 
\special{pn 8}%
\special{sh 0.600}%
\special{ar 376 376 26 26  0.0000000 6.2831853}%
% LINE 2 0 3 0
% 2 375 575 550 575
% 
\special{pn 8}%
\special{pa 376 576}%
\special{pa 550 576}%
\special{fp}%
% LINE 2 0 3 0
% 2 475 575 475 400
% 
\special{pn 8}%
\special{pa 476 576}%
\special{pa 476 400}%
\special{fp}%
% CIRCLE 2 0 0 0
% 4 475 375 475 400 475 400 475 400
% 
\special{pn 8}%
\special{sh 0.600}%
\special{ar 476 376 26 26  0.0000000 6.2831853}%
% LINE 2 0 3 0
% 4 700 575 750 575 750 575 750 575
% 
\special{pn 8}%
\special{pa 700 576}%
\special{pa 750 576}%
\special{fp}%
\special{pa 750 576}%
\special{pa 750 576}%
\special{fp}%
% CIRCLE 2 0 3 0
% 4 675 575 675 600 675 600 675 600
% 
\special{pn 8}%
\special{ar 676 576 26 26  0.0000000 6.2831853}%
% LINE 2 0 3 0
% 2 800 575 850 575
% 
\special{pn 8}%
\special{pa 800 576}%
\special{pa 850 576}%
\special{fp}%
% LINE 2 0 3 0
% 2 875 575 875 400
% 
\special{pn 8}%
\special{pa 876 576}%
\special{pa 876 400}%
\special{fp}%
% CIRCLE 2 0 0 0
% 4 875 375 875 400 875 400 875 400
% 
\special{pn 8}%
\special{sh 0.600}%
\special{ar 876 376 26 26  0.0000000 6.2831853}%
% LINE 2 0 3 0
% 2 775 575 775 400
% 
\special{pn 8}%
\special{pa 776 576}%
\special{pa 776 400}%
\special{fp}%
% CIRCLE 2 0 0 0
% 4 775 375 775 400 775 400 775 400
% 
\special{pn 8}%
\special{sh 0.600}%
\special{ar 776 376 26 26  0.0000000 6.2831853}%
% CIRCLE 2 0 3 0
% 4 975 575 975 600 975 600 975 600
% 
\special{pn 8}%
\special{ar 976 576 26 26  0.0000000 6.2831853}%
% CIRCLE 2 0 3 0
% 4 575 575 575 600 575 600 575 600
% 
\special{pn 8}%
\special{ar 576 576 26 26  0.0000000 6.2831853}%
% LINE 2 0 3 0
% 2 900 575 950 575
% 
\special{pn 8}%
\special{pa 900 576}%
\special{pa 950 576}%
\special{fp}%
% LINE 2 0 3 0
% 2 1000 575 1050 575
% 
\special{pn 8}%
\special{pa 1000 576}%
\special{pa 1050 576}%
\special{fp}%
% LINE 2 0 3 0
% 2 750 575 800 575
% 
\special{pn 8}%
\special{pa 750 576}%
\special{pa 800 576}%
\special{fp}%
% LINE 2 0 3 0
% 2 600 575 650 575
% 
\special{pn 8}%
\special{pa 600 576}%
\special{pa 650 576}%
\special{fp}%
% LINE 2 0 3 0
% 2 850 575 900 575
% 
\special{pn 8}%
\special{pa 850 576}%
\special{pa 900 576}%
\special{fp}%
\end{picture}%

 & \\
(D)					&  \\\hline
					&  \\
%%%%\input{1101010s.tex}
%WinTpicVersion3.08
\unitlength 0.1in
\begin{picture}(  7.0000,  2.5000)(  3.5000, -6.0000)
% LINE 2 0 3 0
% 2 375 575 375 400
% 
\special{pn 8}%
\special{pa 376 576}%
\special{pa 376 400}%
\special{fp}%
% CIRCLE 2 0 0 0
% 4 375 375 375 400 375 400 375 400
% 
\special{pn 8}%
\special{sh 0.600}%
\special{ar 376 376 26 26  0.0000000 6.2831853}%
% LINE 2 0 3 0
% 2 375 575 550 575
% 
\special{pn 8}%
\special{pa 376 576}%
\special{pa 550 576}%
\special{fp}%
% LINE 2 0 3 0
% 2 475 575 475 400
% 
\special{pn 8}%
\special{pa 476 576}%
\special{pa 476 400}%
\special{fp}%
% CIRCLE 2 0 0 0
% 4 475 375 475 400 475 400 475 400
% 
\special{pn 8}%
\special{sh 0.600}%
\special{ar 476 376 26 26  0.0000000 6.2831853}%
% LINE 2 0 3 0
% 4 700 575 750 575 750 575 750 575
% 
\special{pn 8}%
\special{pa 700 576}%
\special{pa 750 576}%
\special{fp}%
\special{pa 750 576}%
\special{pa 750 576}%
\special{fp}%
% CIRCLE 2 0 3 0
% 4 775 575 775 600 775 600 775 600
% 
\special{pn 8}%
\special{ar 776 576 26 26  0.0000000 6.2831853}%
% LINE 2 0 3 0
% 2 800 575 850 575
% 
\special{pn 8}%
\special{pa 800 576}%
\special{pa 850 576}%
\special{fp}%
% LINE 2 0 3 0
% 2 875 575 875 400
% 
\special{pn 8}%
\special{pa 876 576}%
\special{pa 876 400}%
\special{fp}%
% CIRCLE 2 0 0 0
% 4 875 375 875 400 875 400 875 400
% 
\special{pn 8}%
\special{sh 0.600}%
\special{ar 876 376 26 26  0.0000000 6.2831853}%
% LINE 2 0 3 0
% 2 675 575 675 400
% 
\special{pn 8}%
\special{pa 676 576}%
\special{pa 676 400}%
\special{fp}%
% CIRCLE 2 0 0 0
% 4 675 375 675 400 675 400 675 400
% 
\special{pn 8}%
\special{sh 0.600}%
\special{ar 676 376 26 26  0.0000000 6.2831853}%
% CIRCLE 2 0 3 0
% 4 975 575 975 600 975 600 975 600
% 
\special{pn 8}%
\special{ar 976 576 26 26  0.0000000 6.2831853}%
% CIRCLE 2 0 3 0
% 4 575 575 575 600 575 600 575 600
% 
\special{pn 8}%
\special{ar 576 576 26 26  0.0000000 6.2831853}%
% LINE 2 0 3 0
% 2 900 575 950 575
% 
\special{pn 8}%
\special{pa 900 576}%
\special{pa 950 576}%
\special{fp}%
% LINE 2 0 3 0
% 2 1000 575 1050 575
% 
\special{pn 8}%
\special{pa 1000 576}%
\special{pa 1050 576}%
\special{fp}%
% LINE 2 0 3 0
% 2 600 575 700 575
% 
\special{pn 8}%
\special{pa 600 576}%
\special{pa 700 576}%
\special{fp}%
% LINE 2 0 3 0
% 2 850 575 900 575
% 
\special{pn 8}%
\special{pa 850 576}%
\special{pa 900 576}%
\special{fp}%
\end{picture}%

 & 0.1547\\
(E)					&  \\\hline
\end{tabular}
\end{center}
\end{table}
In this case, there are 5 essential patterns
(A) to (E).
We can see that the best pattern is (E).
We can also see that $J(\M)$ depends on the maximal number
of the {\it consecutive zeros} in $\M$
(we here call this the consecutive number).

These results shows that
the optimal value of $J(\M)$ depends on the consecutive number
and not on the number of retained samples.
By this observation, we can make a hypothesis that 
the optimal decimation pattern $\M$ is the pattern
in which the zeros are the least consecutive. 
In other words,  the most uniformly distributed pattern is the best.
In view of this, the block decimation introduced in \cite{NayBarSmi93} cannot be optimal.
%%% R2-1-4
Note that the hypothesis does not detract from the merit of nonuniform decimation;
if the ratio $M/N$ is non integer, nonuniform decimation is inevitable.
%%%

\section{Conclusion}
\label{sec:conclusion}
We have proposed an interpolation method of nonuniform decimation
for non-band-limited signals.
To design the interpolation system,
we adopt the $H^\infty$ norm of the error system.
We have shown that the optimization can be efficiently executed
by numerical computation.
We have also considered designing decimation pattern
with the $H^\infty$ optimal performance index.
Design examples have shown the effectiveness of the
present method.
A theoretical proof for the hypothesis given in Subsection \ref{subsec:ex-pattern}
%that the most uniformly distributed decimation pattern is the best
remains an open question.
%\bibliographystyle{IEEEtran}% bib style
%\bibliography{abrv,jcmsi_noy}% your bib database

\end{document}